\documentclass[a4paper,11pt]{article}
\pdfoutput=1 

\usepackage{jheppub} 

\usepackage[T1]{fontenc} 

\usepackage{amsmath,amssymb}
\pdfoutput=1
\allowdisplaybreaks 

\def \beq{\begin{equation}}
\def \eeq{\end{equation}}
\def \bea{\begin{eqnarray}}
\def \eea{\end{eqnarray}}

\def\bm#1{\mbox{\boldmath$#1$\unboldmath}} 

\title{Simplified dark matter models with two Higgs doublets:  I. Pseudoscalar mediators}

\author[1]{Martin Bauer,}
\author[2,3]{Ulrich Haisch}
\author[4]{and Felix Kahlhoefer}

\affiliation[1]{Institut f{\"u}r Theoretische Physik, Universit{\"a}t Heidelberg, \\ 
Philosophenweg 16, 69120 Heidelberg, Germany}   

\affiliation[2]{Rudolf Peierls Centre for Theoretical Physics,
   University of Oxford, \\ OX1 3NP Oxford, United Kingdom}  
    
\affiliation[3]{CERN, Theoretical Physics Department, \\ CH-1211 Geneva 23, Switzerland}

\affiliation[4]{DESY, Notkestra\ss e 85, D-22607 Hamburg, Germany} 

\emailAdd{bauer@thphys.uni-heidelberg.de, ulrich.haisch@physics.ox.ac.uk,  felix.kahlhoefer@desy.de}

\abstract{We study a new class of renormalisable simplified models for dark matter searches at the LHC that are based on two Higgs doublet models with an additional pseudoscalar mediator. In contrast to the spin-0 simplified models employed in analyses of Run~I data these models are self-consistent, unitary and bounds from Higgs physics typically pose no constraints. Predictions for various missing transverse energy ($E_{T, \rm miss}$) searches are discussed and the reach of the 13~TeV LHC is explored. It is found that the proposed models provide a rich spectrum of complementary observables that lead to non-trivial constraints. We emphasise in this context the sensitivity of the $t\bar t + E_{T, \rm miss}$,  mono-$Z$ and mono-Higgs channels, which yield stronger limits than mono-jet searches in large parts of the parameter space. Constraints from spin-0 resonance searches, electroweak precision measurements and flavour observables are also derived and shown to provide further important handles to constraint and to test the considered dark matter models.}

\preprint{CERN-TH-2017-011, DESY-17-010}

\begin{document} 

\maketitle

\flushbottom

\section{Introduction}
\label{sec:introduction}

Simplified models of dark matter (DM) and a single mediator overcome many of the shortcomings of DM effective field theories, but remain general enough to represent a large class of popular theories of DM (see the reviews \cite{Abdallah:2014hon,Abdallah:2015ter, Abercrombie:2015wmb} for a complete list of references). In particular, including contributions from on-shell production of the mediators allows to capture the full kinematics of DM production at colliders, making meaningful comparisons with bounds from direct and indirect detection experiments possible. 

In simplified DM models the  interactions between the mediators and the Standard Model~(SM) fermions are usually written as gauge or Yukawa couplings of mass dimension four. In many cases these interactions are however only apparently renormalisable, because in a full $SU(2)_L \times U(1)_Y$ gauge-invariant theory they in fact arise from  higher-dimensional operators or  they signal the presence of additional particles or couplings that are needed to restore gauge invariance \cite{Chala:2015ama,Bell:2015sza,Kahlhoefer:2015bea,Bell:2015rdw,Haisch:2016usn,Englert:2016joy}. These features can lead to parameter regions which are theoretically inaccessible or to misleading/unphysical predictions often related to unitarity violation. Models in which the mediators mix with the SM bosons avoid such inconsistencies. The existing LEP and LHC measurements of the $Z$-boson and Higgs-boson properties however  severely restrict the corresponding mixing angles, and as a result classic $E_{T, \rm miss}$ searches like mono-jets are typically not the leading collider constraints in  this class of simplified DM models \cite{Kahlhoefer:2015bea,Duerr:2016tmh,Bauer:2016gys}.

In this article, we study a new class of simplified DM models for spin-0 mediators based on two Higgs doublet models (THDMs), which are an essential ingredient of many well-motivated theories beyond the SM. In contrast to inert THDMs, where the DM particle is the lightest neutral component of the second Higgs doublet and is stabilised by an ad-hoc~$Z_2$~symmetry \cite{Deshpande:1977rw,Ma:2006km,Barbieri:2006dq,LopezHonorez:2006gr}, our focus is on the case where the DM candidate is a SM singlet fermion. To couple the DM particle to the SM, we introduce a new spin-0 mediator, which mixes dominantly with the scalar or pseudoscalar partners of the SM Higgs. In this way constraints from Higgs signal strength measurements~\cite{ATLAS-CONF-2015-044} can be satisfied and one obtains a framework in which all operators are gauge invariant and renormalisable.

In what follows we will explore the phenomenology of pseudoscalar mediators, while  scalar portals  will be discussed in detail in an accompanying paper \cite{partII} (see also~\cite{Bell:2016ekl}). Pseudoscalar mediators have the obvious advantage of avoiding constraints from DM direct-detection experiments, so that the observed DM relic abundance can be reproduced in large regions of parameter space and LHC searches are particularly relevant to test these models. Similar investigations of THDM plus pseudoscalar simplified DM  models have been presented  in~\cite{Ipek:2014gua,No:2015xqa,Goncalves:2016iyg}. Whenever indicated we will highlight the similarities and differences between these and our work.

\begin{figure}[!t]
\begin{center}
\includegraphics[height=0.2\textwidth]{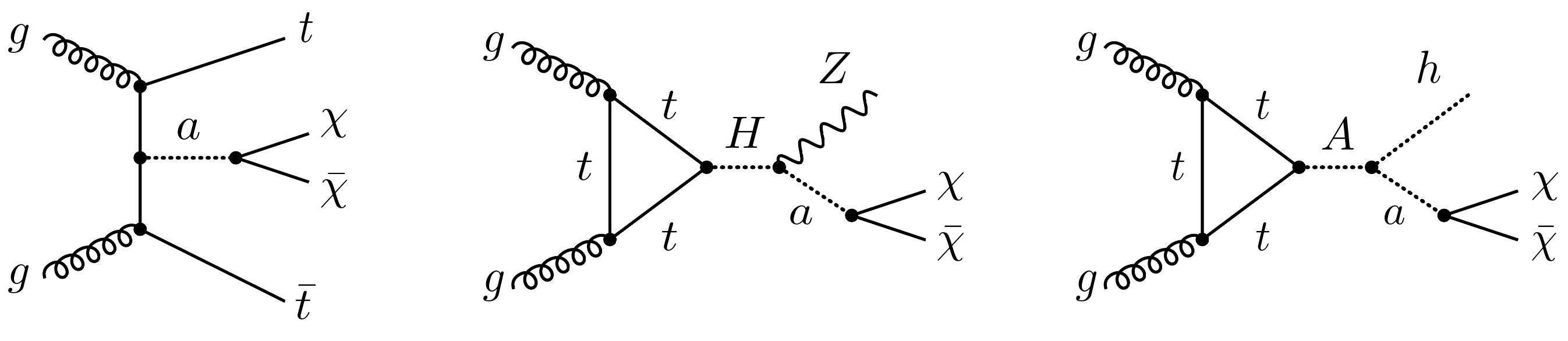}  
\vspace{0mm}
\caption{\label{fig:assortedmet} Assorted diagrams that give rise to a $t \bar t + E_{T, \rm miss}$ (left),  $Z+ E_{T, \rm miss}$ (middle) and  $h + E_{T, \rm miss}$~(right) signal in the  simplified pseudoscalar model considered in our work. The exchanged spin-0 particles are of scalar ($H$) or pseudoscalar ($a, A$) type. Further Feynman graphs that contribute to the different mono-$X$ channels  can be found in Figures~\ref{fig:jDMDM} to \ref{fig:WDMDM}.}
\end{center}
\end{figure}

The mono-$X$ phenomenology of the considered simplified pseudoscalar models turns out to be surprisingly rich.  We examine the constraints from searches for $j+ E_{T, \rm miss}$~\cite{Aaboud:2016tnv,CMS:2016pod}, $t\bar t/b \bar b + E_{T, \rm miss}$~\cite{ATLAS:2016ljb,CMS:2016mxc,ATLAS-CONF-2016-086,CMS:2016uxr},  $Z+E_{T, \rm miss}$~\cite{ATLAS:2016bza,CMS:2016hmx, Sirunyan:2017onm}, $h+E_{T, \rm miss}$~\cite{Aaboud:2016obm,CMS:2016mjh,ATLAS-CONF-2016-011,CMS:2016xok}  and $W+E_{T, \rm miss}$~\cite{Aaboud:2016zkn,Khachatryan:2016jww} and present projections for the 13~TeV LHC. In particular, we provide benchmark scenarios that are consistent with bounds from electroweak (EW) precision, flavour and Higgs observables including invisible decays~\cite{Aad:2015pla,Khachatryan:2016whc}. For the simplified pseudoscalar model recommended by the ATLAS/CMS DM Forum (DMF)~\cite{Abercrombie:2015wmb} constraints from mono-jet searches dominate throughout the parameter space~\cite{Haisch:2015ioa}, whereas for the model considered here $t\bar t + E_{T, \rm miss}$,  mono-$Z$ and mono-Higgs searches yield competitive bounds and often provide the leading constraints. See Figure~\ref{fig:assortedmet} for an illustration of the various $E_{T, \rm miss}$ processes that are of most interest in our simplified model. This complementarity of different searches is the result of the consistent treatment of the scalar sector, inducing gauge and trilinear scalar couplings of the mediator beyond the ones present in the DMF pseudoscalar model. 

It is particularly appealing that the $Z+ E_{T, \rm miss}$ and $h + E_{T, \rm miss}$ signatures are strongest in the theoretically best motivated region of parameter space, where the couplings of the light Higgs  are SM-like.   In this region of parameter space, couplings of the new scalar states to SM gauge bosons are strongly suppressed and play no role in the phenomenology, leading to gluon-fusion dominated production and a very predictive pattern of branching ratios. In consequence, a complementary search strategy can be advised, with the exciting possibility to observe DM simultaneously in a number of different channels, some of which are not limited by systematic errors and can be improved by statistics even beyond~$300 \,  {\rm fb}^{-1}$ of luminosity. The importance of di-top resonance searches~\cite{ATLAS:2016pyq,HaischTeVPA16} to probe  neutral spin-0 states with masses above the $t \bar t$ threshold is also stressed, and it is  pointed out that for model realisations with a light scalar partner of the SM Higgs, di-tau resonance searches should provide relevant constraints in the near future. We finally comment on the impact of bottom-quark ($b \bar b)$ initiated production.

This paper is structured as follows. In Section~\ref{sec:THDMP} we describe the class of simplified DM models that we will study throughout our work, while Section~\ref{sec:anatomy} contains a comprehensive review of the non-$E_{T, \rm miss}$ constraints that have to be satisfied in order to make a given model realisation phenomenologically viable. The partial decay widths and the branching ratios of the spin-0 particles arising in the considered simplified DM models are studied in Section~\ref{sec:widths}. The most important features of the resulting $E_{T, \rm miss}$  phenomenology are described in~Section~\ref{sec:METsignals}. In~Section~\ref{sec:numerics} we finally present the numerical results of our analyses providing summary plots of the mono-$X$ constraints for several benchmark scenarios. The result-oriented reader might want to skip directly to this section. Our conclusions and a brief outlook are given in~Section~\ref{sec:conclusions}. 

\section{THDM plus pseudoscalar extensions}
\label{sec:THDMP}

In this section we describe the structure of the simplified DM model with a pseudoscalar mediator. We start with the scalar potential and then consider the Yukawa sector. In both cases we will point out which are the new parameters corresponding to the interactions in question.

\subsection{Scalar potential}
\label{sec:scalarpotential}

The tree-level THDM scalar potential that we will consider throughout this paper is given by the following expression (see for example~\cite{Gunion:1989we,Branco:2011iw})
\beq \label{eq:VH}
\begin{split}
V_{H} & = \mu_1 H_1^\dagger H_1 + \mu_2 H_2^\dagger H_2 + \left ( \mu_3  H_1^\dagger H_2 + {\rm h.c.} \right ) + \lambda_1  \hspace{0.25mm} \big ( H_1^\dagger H_1  \big )^2  + \lambda_2  \hspace{0.25mm} \big ( H_2^\dagger H_2 \big  )^2 \\[2mm]
& \phantom{xx} +  \lambda_3 \hspace{0.25mm} \big ( H_1^\dagger H_1  \big ) \big ( H_2^\dagger H_2  \big ) + \lambda_4  \hspace{0.25mm} \big ( H_1^\dagger H_2  \big ) \big ( H_2^\dagger H_1  \big ) + \left [ \lambda_5   \hspace{0.25mm} \big ( H_1^\dagger H_2 \big )^2 + {\rm h.c.} \right ]  \,.
\end{split} 
\eeq
Here we have imposed a $Z_2$ symmetry under which $H_1 \to H_1$ and $H_2 \to -H_2$ to suppress flavour-changing neutral currents (FCNCs), but allowed for this discrete symmetry to be softly broken by the term $\mu_3  H_1^\dagger H_2 + {\rm h.c.}$ The vacuum expectation values (VEVs) of the Higgs doublets are given by $\langle H_i \rangle = (0,v_i/\sqrt{2})^T$ with $v = \sqrt{v_1^2 + v_2^2} \simeq 246 \, {\rm GeV}$ and we  define $\tan \beta = v_2/v_1$. To avoid possible issues with electric dipole moments, we assume that the mass-squared terms $\mu_j$, the quartic couplings~$\lambda_k$ and the VEVs are all real and as a result  the scalar potential as given in (\ref{eq:VH}) is CP conserving. The three physical neutral Higgses that emerge from $V_H$ are in such a case both mass and CP eigenstates.

The most economic way to couple fermionic DM to the SM through pseudoscalar exchange is by mixing a CP-odd mediator $P$ with the CP-odd Higgs that arises from (\ref{eq:VH}). This can be achieved by considering the following interaction terms 
\beq \label{eq:VP}
\begin{split}
V_{P}  =  \frac{1}{2} \hspace{0.5mm} m_P^2  P^2 +  P \left ( i \hspace{0.1mm} b_P  \hspace{0.1mm}  H_1^\dagger H_2 + {\rm h.c.} \right ) + P^2 \left (  \lambda_{P1}  \hspace{0.1mm}  H_1^\dagger H_1 +   \lambda_{P2}  \hspace{0.1mm}  H_2^\dagger H_2 \right )  \,,
\end{split} 
\eeq
where $m_P$ and $b_P$ are parameters with dimensions of mass. We assume that $V_{P}$ does not break CP and thus take $b_P$  to be real in the following. In this case $P$ does not develop a VEV and remains a  pure CP eigenstate. Nevertheless, this term does lead to a soft breaking of the $Z_2$ symmetry. Notice that compared to~\cite{Ipek:2014gua,No:2015xqa,Goncalves:2016iyg} which include only the trilinear portal coupling $b_P$, we also allow for quartic portal interactions proportional to~$\lambda_{P1}$ and $\lambda_{P2}$. A~quartic self-coupling of the form $P^4$ has instead not been included in (\ref{eq:VP}), as it does not lead to any relevant effect in the observables studied in our paper.  

The interactions in the scalar potential  (\ref{eq:VH}) mix the neutral CP-even weak eigenstates and we denote the corresponding mixing angle by $\alpha$. The portal coupling $b_P$ appearing in~(\ref{eq:VP}) instead mixes the two neutral CP-odd weak eigenstates with $\theta$ representing the associated mixing angle. The resulting  CP-even mass eigenstates will be denoted by $h$ and~$H$, while in the~CP-odd sector the states will be called $A$ and $a$, where $a$  denotes the extra degree of freedom not present in THDMs. The scalar spectrum also contains two charged mass eigenstates $H^\pm$ of identical mass. 

Diagonalising the mass-squared matrices of the scalar states  leads to relations between the fundamental parameters entering $V_H$ and $V_P$. These relations allow to trade the parameters $m_P$, $\mu_1$, $\mu_2$, $\mu_3$, $b_P$, $\lambda_1$, $\lambda_2$, $\lambda_4$, $\lambda_5$ for sines and cosines of mixing angles, VEVs and the masses of the physical Higgses. This procedure  ensures in addition  that the scalar potential is positive definite and that the vacuum solution is an absolute minimum. In the broken EW phase the physics of (\ref{eq:VH}) and (\ref{eq:VP}) is hence fully captured by the angles~$\alpha$, $\beta$, $\theta$,  the EW VEV $v$, the quartic couplings $\lambda_3$, $\lambda_{P1}$, $\lambda_{P2}$ and the masses $M_h$, $M_H$, $M_A$, $M_{H^\pm}$, $M_a$. We will use these parameters as input in our analysis. 

\subsection{Yukawa sector}
\label{sec:yukawasector}

The couplings between the scalars and the SM fermions are restricted by the stringent experimental limits on flavour observables. A necessary and sufficient condition to avoid~FCNCs associated to neutral Higgs tree-level exchange is that not more than one of the Higgs doublets couples to fermions of a given charge \cite{Glashow:1976nt,Paschos:1976ay}. This so-called natural flavour conservation hypothesis is automatically enforced by the aforementioned $Z_2$ symmetry acting on the doublets, if the right-handed fermion singlets transform accordingly. The Yukawa couplings are explicitly given by
\beq \label{eq:LY}
{\cal L}_{Y} = - \sum_{i=1,2} \left ( \bar Q Y_u^i \tilde H_i u_R  + \bar Q Y_d^i H_i d_R   + \bar L Y_\ell^i H_i \ell_R  + {\rm h.c.}  \right ) \,.
\eeq
Here $Y_f^i$ are Yukawa matrices acting on the three fermion generations and we have suppressed flavour indices, $Q$ and $L$ are left-handed quark and lepton doublets, while $u_R$, $d_R$ and $\ell_R$ are right-handed up-type quark, down-type quark and charged lepton singlets, respectively. Finally, $\tilde H_i = \epsilon H_i^\ast$ with $\epsilon$ denoting the  two-dimensional antisymmetric tensor. The natural flavour conservation hypothesis can be satisfied by four discrete assignments, where by convention up-type quarks are always taken to couple to $H_2$:
\beq \label{eq:yukawatypes}
\begin{split}
Y_u^1 & = Y_d^1 = Y_\ell^1 =0 \,, \quad (\text{type I}) \,, \\[2mm]
Y_u^1 & = Y_d^2 = Y_\ell^2 =0 \,, \quad (\text{type II}) \,, \\[2mm]
Y_u^1 & = Y_d^1 = Y_\ell^2 =0\,, \quad (\text{type III}) \,, \\[2mm]
Y_u^1 & = Y_d^2 = Y_\ell^1 =0\,, \quad (\text{type IV}) \,. \\[2mm]
\end{split}
\eeq
The dependence of our results on the choice of the Yukawa sector will be discussed in some detail in the next section. 

Taking DM to be a Dirac fermion  $\chi$ a separate $Z_2$ symmetry under which $\chi \to -\chi$ can be used to forbid a coupling of the form $\bar L \tilde H_1 \chi_R +{\rm h.c.}$ At the level of renormalisable operators this leaves 
\beq \label{eq:Lx}
{\cal L}_\chi = - i \hspace{0.25mm} y_\chi P \hspace{0.25mm} \bar \chi \hspace{0.25mm} \gamma_5 \hspace{0.1mm} \chi \,,
\eeq
as the only possibility to couple the pseudoscalar mediator $P$ to DM. In order to not violate~CP we require the dark sector Yukawa coupling $y_\chi$ to be real. The parameter $y_\chi$ and the DM mass $m_\chi$ are further input parameters in our analysis. 

\section{Anatomy of the parameter space}
\label{sec:anatomy}

In this section we examine the anatomy of the parameter space of the model introduced above and discuss a number of important simplifications. We briefly explain the alignment/decoupling limit and describe the dependence of the predictions on the choice of Yukawa sector. The constraints on the mixing angles, quartic couplings and Higgs masses from spin-0 resonance searches,  flavour physics,  EW precision measurements, perturbativity and unitarity are also elucidated. 

\subsection{Alignment/decoupling limit}
\label{sec:adlimit}

After EW symmetry breaking  the kinetic terms of the Higgs fields $H_i$  lead to interactions between the CP-even mass eigenstates and the massive EW gauge bosons. These interactions take the form 
\beq \label{eq:hHVV}
{\cal L} \supset  \Big ( \sin \left (\beta - \alpha \right )  h + \cos \left (\beta - \alpha \right ) H \Big )   \left ( \frac{2 M_W^2}{v} \hspace{0.25mm} W_\mu^+ W^{- \mu}  + \frac{M_Z^2}{v} \hspace{0.25mm} Z_\mu Z^\mu  \right ) \,.
\eeq
In order to simplify the further analysis, we concentrate on the well-motivated alignment/decoupling limit of the~THDM where $\alpha = \beta - \pi/2$. In this case $\sin \left ( \beta - \alpha \right ) = 1$ meaning that  the field~$h$ has SM-like EW~gauge boson couplings. It can therefore be identified with the boson of mass~$M_h \simeq 125 \, {\rm GeV}$ discovered at the LHC and the constraints from the Run I combination of the ATLAS and CMS measurements of the Higgs boson production and decay rates to SM final states \cite{ATLAS-CONF-2015-044} are readily fulfilled. Notice that in the alignment/decoupling limit the scalar~$H$ does not interact with $W$-boson or $Z$-boson pairs at tree level because in this limit one has $\cos \left ( \beta - \alpha \right ) = 0$.

\subsection{Yukawa assignments}
\label{sec:yukawas}

Working in the alignment/decoupling limit the fermion-scalar interactions most relevant for the further discussion  are given by 
\beq \label{eq:fermioninteractions}
\begin{split}
{\cal L} \supset &  - \frac{y_t}{\sqrt{2}} \,  \bar t \, \Big [ \hspace{0.25mm} h + \xi_f^{\rm M} H  - i   \hspace{0.5mm} \xi_f^{\rm M}  \, \big ( \hspace{-0.1mm} \cos \theta \, A - \sin \theta \, a \big )  \gamma_5 \Big ]  \,  t \\[2mm]
& -\sum_{f=b,\tau} \frac{y_f}{\sqrt{2}} \,  \bar f \, \Big [ \hspace{0.25mm} h + \xi_f^{\rm M} H  + i   \hspace{0.5mm} \xi_f^{\rm M}  \, \big ( \hspace{-0.1mm} \cos \theta \, A - \sin \theta \, a \big )  \gamma_5 \Big ]  \, f   \\[2mm]
& -  \frac{y_t}{\sqrt{2}} \hspace{0.1mm} V_{tb} \, \xi_t^{\rm M}   H^+ \hspace{0.25mm} \bar t_R \hspace{0.25mm} b_L +  \frac{y_b}{\sqrt{2}} \hspace{0.25mm}  V_{tb}  \, \xi_b^{\rm M}   H^+ \hspace{0.25mm} \bar t_L \hspace{0.25mm} b_R + {\rm h.c.}  \\[2mm]
& - i  \hspace{0.25mm} y_\chi \, \Big ( \hspace{-0.25mm} \sin \theta \, A + \cos \theta \, a \Big ) \,  \bar \chi \hspace{0.25mm}  \gamma_5 \hspace{0.1mm} \chi \,,
\end{split}
\eeq
where $y_f = \sqrt{2} m_f/v$ denote the SM Yukawa couplings and $V_{ij}$ are the elements of the  Cabibbo-Kobayashi-Maskawa (CKM) matrix. The couplings $\xi_f^{\rm M}$ encode the dependence on the choice of Yukawa sector (\ref{eq:yukawatypes}). In terms of $\tan \beta$ one has 
\beq \label{eq:xicouplings}
\begin{split}
& \hspace{13mm} \xi_t^{\rm I}   =  \xi_b^{\rm I} = \xi_\tau^{\rm I} = -\cot \beta \,, \quad (\text{type I}) \,, \\[2mm]
& \hspace{0mm}  \xi_t^{\rm II}  = -\cot \beta \,, \qquad \xi_b^{\rm II}  =  \xi_\tau^{\rm II}  = \tan \beta \,, \quad (\text{type II}) \,, \\[2mm]
& \xi_t^{\rm III}  = \xi_b^{\rm III}  = -\cot \beta \,, \qquad \xi_\tau^{\rm III}  = \tan \beta \,, \quad (\text{type III}) \,, \\[2mm]
&\hspace{1mm}  \xi_t^{\rm IV}  =  \xi_\tau^{\rm IV}  = -\cot \beta \,, \qquad \xi_b^{\rm IV}  = \tan \beta \,, \quad (\text{type IV}) \,.
\end{split}
\eeq

Since the production of the pseudoscalar mediator $a$ as well as $pp \to h, H, A$ is driven by top-quark loops that enter  the gluon-fusion ($gg$) channel at the LHC (see for instance~\cite{Haisch:2012kf} for a discussion in the context of $E_{T, {\rm miss}}$ searches) we will  in the following focus on the region of small $\tan \beta$.  In this limit  the couplings of $H,A,a$  to down-type quarks and charged leptons in~(\ref{eq:fermioninteractions})  are strongly Yukawa suppressed irrespectively of the chosen Yukawa assignment~(\ref{eq:xicouplings}). As a result existing bounds on the neutral scalar masses from flavour observables such as $B_s \to \mu^+ \mu^-$ that are known to receive $\tan \beta$ enhanced corrections~\cite{Bobeth:2001sq} are within experimental limits \cite{CMS:2014xfa} even for a light scalar spectrum.

\subsection{Di-tau searches}
\label{sec:ditau}

In order to understand whether the existing LHC searches for heavy neutral Higgses in fermionic final states such as $f  \bar f = \tau^+ \tau^-, b \bar b$ pose constraints on the low $\tan \beta$ region of our simplified model, it is important to  realise that while the pseudoscalars~$A$ and $a$ couple both to DM, the heavy scalar~$H$ does not, as can be seen from~(\ref{eq:fermioninteractions}). If the channels $A/a \to \chi \bar \chi$ are open, the discovery potential for $H \to f \bar f$ is therefore generically larger than that for the corresponding pseudoscalar modes. In fact, the constraints from $pp \to H \to f \bar{f}$ are most stringent for model realisations with $M_H < 2 m_t$ and $M_a > \text{max} \left (M_H - M_Z, M_H/2 \right )$, so that the decays $H \to t\bar{t}$, $H \to a a$ and $H \to a Z$ are kinematically forbidden and in consequence $H$ is forced to decay to light SM fermions (see Section~\ref{sec:heavierscalar}). 

The typical restrictions that result from  LHC searches for heavy scalars can be illustrated by considering $M_H = 300 \, {\rm GeV}$ and  employing the 95\% confidence level (CL) limit $\sigma \left ( p p \to H \right ) {\rm BR} \left ( H \to \tau^+ \tau^- \right ) < 0.4 \, {\rm pb}$~\cite{ATLAS-CONF-2016-085, CMS:2016rjp} that is based on $13 \, {\rm fb}^{-1}$ of 13 TeV  data. Using the next-to-next-to-next-to-leading order results~\cite{Anastasiou:2016hlm} for inclusive~$H$ production in gluon fusion, we then find that the current di-tau searches only exclude a narrow sliver of parameters in the $M_a$--$\hspace{0.5mm} \tan \beta$ plane with $0.55 \lesssim \tan \beta \lesssim 0.65$ and $M_a \gtrsim 210 \, {\rm GeV}$ in the case of a Yukawa sector of type II.  A reduction of the quoted upper limit on the production cross section times branching ratio to $0.2 \, {\rm pb}$ would however improve the range of excluded~$\tan \beta$ values to $0.3 \lesssim \tan \beta \lesssim 1.2$. As we will see in Section~\ref{sec:wetdream}, such a constraint would be very valuable because probing models with $\tan \beta = {\cal O} (1)$ and $M_H \simeq M_a \simeq 300 \, {\rm GeV}$ turns out to be   difficult  by other means. 

\subsection{Di-top searches}
\label{sec:ditop}

Heavy scalar  and pseudoscalar bosons decaying dominantly into  top-quark pairs can be searched for by studying the resulting $t \bar t$ invariant mass spectra $m_{t \bar t}$. In contrast to di-top searches for spin-1 or spin-2 states, a peak in the $m_{t \bar t}$ distribution that one generically expects in the narrow-width approximation (NWA) is however not the only signature of a spin-0 resonance in this case. Indeed, the $gg \to H/A$ signal will interfere with the QCD $t \bar t$ background which at the LHC is mainly generated by the gluon-fusion channel $gg \to t \bar t$. The signal-background interference will depend on the CP nature of the intermediate spin-0 boson, its mass and its total decay width. The observed interference pattern can be either constructive or destructive, leading to a rather complex signature with a peak-dip structure in the $m_{t \bar t}$ spectrum~\cite{Dicus:1994bm,Frederix:2007gi}. The $pp \to H/A \to t \bar t$ channel provides hence an interesting but challenging opportunity for hadron colliders to search for additional spin-0 bosons (see for instance \cite{Djouadi:2015jea,Craig:2015jba} for recent phenomenological discussions). 

The first LHC analysis that takes into account interference effects between the signal process $gg \to H/A \to t \bar t$ and the SM background $gg \to  t \bar t$ is the ATLAS search \cite{ATLAS:2016pyq}. It is based on $20.3 \, {\rm fb}^{-1}$ of 8 TeV LHC data and considers the $m_{t \bar t}$ spectrum in final states with a single charged lepton (electron or muon), large $E_{T, \rm miss}$ and at least four jets. The search results are interpreted in the context of a pure THDM of type II for two different mass points and employ the alignment/decoupling limit,~i.e.~$\sin \left ( \beta - \alpha \right ) = 1$. For a neutral scalar~$H$~(pseudoscalar~$A$) with a mass of $500 \, {\rm GeV}$, the ATLAS analysis excludes the parameter values $\tan \beta < 0.45$ ($\tan \beta < 0.85$) at the 95\%~CL, while for the $750 \, {\rm GeV}$ mass point  no meaningful constraint on $\tan \beta$ can be set. Recasting these limits into bounds on the parameter space of spin-0 simplified DM models is straightforward~\cite{HaischTeVPA16} and we will analyse the resulting restrictions on our model in~Section~\ref{sec:wetdream}. 

\subsection{Flavour constraints}
\label{sec:flavour}

Indirect constraints on the charged Higgs-boson mass $M_{H^\pm}$ arise from $Z \to b \bar b$~\cite{Denner:1991ie,Haisch:2007ia,Freitas:2012sy},  $B \to X_s \gamma$~\cite{Hermann:2012fc,Misiak:2015xwa,Czakon:2015exa} and $B_q$--$\bar B_q$  mixing \cite{Abbott:1979dt,Geng:1988bq,Buras:1989ui,Eberhardt:2013uba} since the latter processes receive corrections from the $H^+ \hspace{0.25mm} \bar t_R \hspace{0.25mm} b_L + {\rm h.c.}$ and $H^+ \hspace{0.25mm} \bar t_L \hspace{0.25mm} b_R + {\rm h.c.}$ terms in~(\ref{eq:fermioninteractions}). We find that $B \to X_s \gamma$ provides the strongest indirect constraint on $M_{H^\pm}$ for small $\tan \beta$ values in models of type~I~and~III at present, while $B_s$--$\bar B_s$ oscillations represent the leading indirect constraint in the other two cases. For  $M_{H^\pm} = 750 \, {\rm GeV}$ we obtain the  bound $\tan \beta \gtrsim 0.8$ from a combination of $B$-meson physics observables irrespective of the choice of the Yukawa sector. A model-independent lower limit of $\tan \beta \gtrsim 0.3$ can also be obtained from the requirement that the top-quark Yukawa coupling remains perturbative~\cite{Branco:2011iw}.  The latest LHC search limits on the charged Higgs mass in the $pp \to tbH^\pm \, (H^\pm \to tb)$ channel \cite{Khachatryan:2015qxa,ATLAS:2016qiq} are satisfied for~$\tan \beta \gtrsim 0.2$ if $M_{H^\pm} = 750 \, {\rm GeV}$ is assumed, and therefore provide no relevant constraint. 

\subsection{EW precision constraints}
\label{sec:EWprecision}
 
A scalar potential with two doublets such as the one introduced in (\ref{eq:VH}) leads  to additional Higgs interactions compared to the SM, which can violate the custodial symmetry present in the SM Higgs sector. It can be shown~\cite{Haber:1992py,Pomarol:1993mu,Gerard:2007kn,Grzadkowski:2010dj,Haber:2010bw} that the tree-level potential $V_H$ is custodially invariant for  $M_A = M_{H^\pm}$ or $M_H = M_{H^\pm}$. Only in these two cases can $H$ or $A$ have a sizeable mass splitting from the rest of the Higgses without being in conflict with EW precision measurements, most importantly~$\Delta \rho$. Since the potential  (\ref{eq:VP}) mixes the  pseudoscalar degree of freedom in~$H_i$ with $P$, in the theory described by $V_H + V_P$ there are however additional sources of custodial symmetry breaking compared to the case of the pure THDM. In the alignment/decoupling limit and taking  $M_A = M_{H^\pm}$, we find that the extended scalar sector gives rise to the following one-loop correction
\beq \label{eq:Deltarho}
\Delta \rho = \frac{1}{(4 \pi)^2} \frac{M_{H^\pm}^2}{v^2} \, \Big  [ 1 + f (M_H, M_a, M_{H^\pm}  ) + f  (M_a, M_H, M_{H^\pm}  )   \Big  ]  \hspace{0.1mm} \sin^2 \theta \,,
\eeq
with 
\beq \label{eq:f3}
f (m_1, m_2, m_3  ) = \frac{m_1^4 \left ( m_2^2 - m_3^2 \right)}{m_3^2 \left (m_1^2 - m_2^2 \right ) \left ( m_1^2 - m_3^2 \right )}  \hspace{0.1mm} \ln \left ( \frac{m_1^2}{m_3^2} \right ) \,.
\eeq
Notice that $\Delta \rho \to 0$ in the limit $\sin \theta \to 0$ in which the two CP-odd weak eigenstates are also mass eigenstates or  if  the scalar mass spectrum is fully degenerate. In the alignment/decoupling limit with $M_H = M_{H^\pm}$,  the custodial symmetry is instead not broken by~$V_H + V_P$  and as a result one has $\Delta \rho = 0$  at the one-loop level. 

From the above discussion it follows that only cases with $M_A = M_{H^\pm}$ are subject to the constraints from the EW precision measurements, while scenarios with $M_H = M_{H^\pm}$ are~not.  In order to derive the resulting constraints in the former case, we employ the~95\%~CL bound 
\beq \label{eq:Deltarhoexp}
\Delta \rho \in [-1.2, 2.4 ] \cdot 10^{-3} \,,
\eeq
which corresponds to the value extracted in~\cite{Olive:2016xmw} from a simultaneous determination of the Peskin-Takeuchi parameters $S$, $T$ and $U$. The fact that (\ref{eq:Deltarho}) is proportional to the product of mass differences $M_{H^\pm} - M_H$ and $M_{H^\pm} - M_a$ as well as $\sin^2 \theta$ implies that the existing EW precision data allow to set stringent bounds on $\sin \theta$ if the relevant mass splittings in the scalar sector are sizeable. Taking for instance $M_{H^\pm} = 750 \, {\rm GeV}$ and $M_{a} = 65 \, {\rm GeV}$, we find  that for $M_H = 500 \, {\rm GeV}$ ($M_H = 300 \, {\rm GeV}$) the inequality $\sin \theta < 0.35$ ($\sin \theta < 0.25$) has to be satisfied in order to be compatible with (\ref{eq:Deltarhoexp}). We will see in~Section~\ref{sec:heavierscalar} that the  restrictions on $\sin \theta$ can have a visible impact on the decay pattern of the scalar~$H$, which in turn affects the mono-$Z$ phenomenology discussed in Section~\ref{sec:wetdream}. 

\subsection{Perturbativity and unitarity}
\label{sec:perturbativityunitarity}
 
Perturbativity \cite{Gunion:2002zf,Barroso:2013awa} and unitarity \cite{Kanemura:1993hm,Akeroyd:2000wc,Ginzburg:2005dt,Grinstein:2015rtl} also put restrictions on the scalar masses and  the magnitudes and signs of the quartic couplings. In our numerical analysis we will restrict our attention to the parameter space that satisfies $M_H, M_A, M_a \leq M_{H^\pm} = {\cal O} ( 1 \, {\rm TeV})$ and always keep $\lambda_3$, $\lambda_{P1}$ and $\lambda_{P2}$ of ${\cal O} (1)$ or below.  For such input parameter choices all constraints discussed in this section are satisfied if $\tan \beta$ is not too far below~1. We also only consider parameters for which the total decay widths of $H$ and $A$ are sufficiently small so that the NWA applies, i.e.~$\Gamma_{i} \lesssim M_i/3$ for $i=H,A$. This requirement sets an upper limit on the mass of the charged Higgs boson that is often  stronger than bounds from perturbativity.

\section{Partial decay widths and branching ratios}
\label{sec:widths}

This section is devoted to the discussion of the partial decay widths and the branching ratios of the spin-0 particles arising in the simplified DM model introduced in Section~\ref{sec:THDMP}. For concreteness we will focus on the alignment/decoupling  limit of the theory.  We will  furthermore pay special attention to the parameter space with a light DM particle, small values of $\tan \beta$ and scalar spectra where the  new pseudoscalar $a$ and the scalar $h$ are the lightest degrees of freedom.
 
\subsection[Lighter pseudoscalar $a$]{Lighter pseudoscalar $\bm{a}$}
\label{sec:lighterpseudoscalar}

As a result of CP conservation the field $a$ has no couplings of the form $aW^+W^-$, $aZZ$ and $ahh$.  In contrast the $ahZ$ vertex is allowed by CP symmetry but vanishes  in the  alignment/decoupling  limit. At tree level the pseudoscalar $a$ can thus only decay  into DM particles and SM fermions. The corresponding partial decay widths are given by
\beq \label{eq:Gammah4tree}
\begin{split}
\Gamma \left ( a \to \chi \bar \chi \right ) & = \frac{y_\chi^2}{8\pi} \hspace{0.25mm} M_a  \hspace{0.25mm}  \beta_{\chi/a} \cos^2 \theta \,, \\[2mm]
\Gamma \left ( a \to f \bar f \right ) & = \frac{N_c^f \big ( \xi_f^{\rm M} \big)^2}{8\pi}  \hspace{0.1mm}  \frac{m_f^2}{v^2}  \hspace{0.25mm}  M_a  \hspace{0.25mm}  \beta_{f/a}  \sin^2 \theta  \,,
\end{split}
\eeq
where  $\beta_{i/a} = \sqrt{1 - \tau_{i/a}}$ is the velocity of the particle $i$ in the  rest frame of the final-state pair and we have defined  $\tau_{i/a}=4 m_i^2/M_a^2$. Furthermore $N_c^f = 3 \, (1)$ denotes the relevant colour factor for quarks (leptons) and the explicit expressions for the couplings $\xi_f^{\rm M}$  can be found in (\ref{eq:xicouplings}). At the loop level the pseudoscalar~$a$ can also decay  to gauge bosons. The largest partial decay width is the one to  gluon pairs. It  takes the form 
\beq \label{eq:Gammah4loop}
\Gamma \left (a \to gg \right ) = \frac{\alpha_s^2}{32 \pi^3 v^2} \, M_a^3 \, \Big | \sum_{q=t,b,c} \xi_q^{\rm M} \hspace{0.25mm} f(\tau_{q/a}) \Big |^2 \, \sin^2 \theta \,,
\eeq
with 
\beq \label{eq:ffgluon}
f (\tau) = \tau \arctan^2 \left ( \frac{1}{\sqrt{\tau -1 }} \right ) \,.
\eeq

\begin{figure}[t!]
\begin{center}
\includegraphics[width=0.425\textwidth]{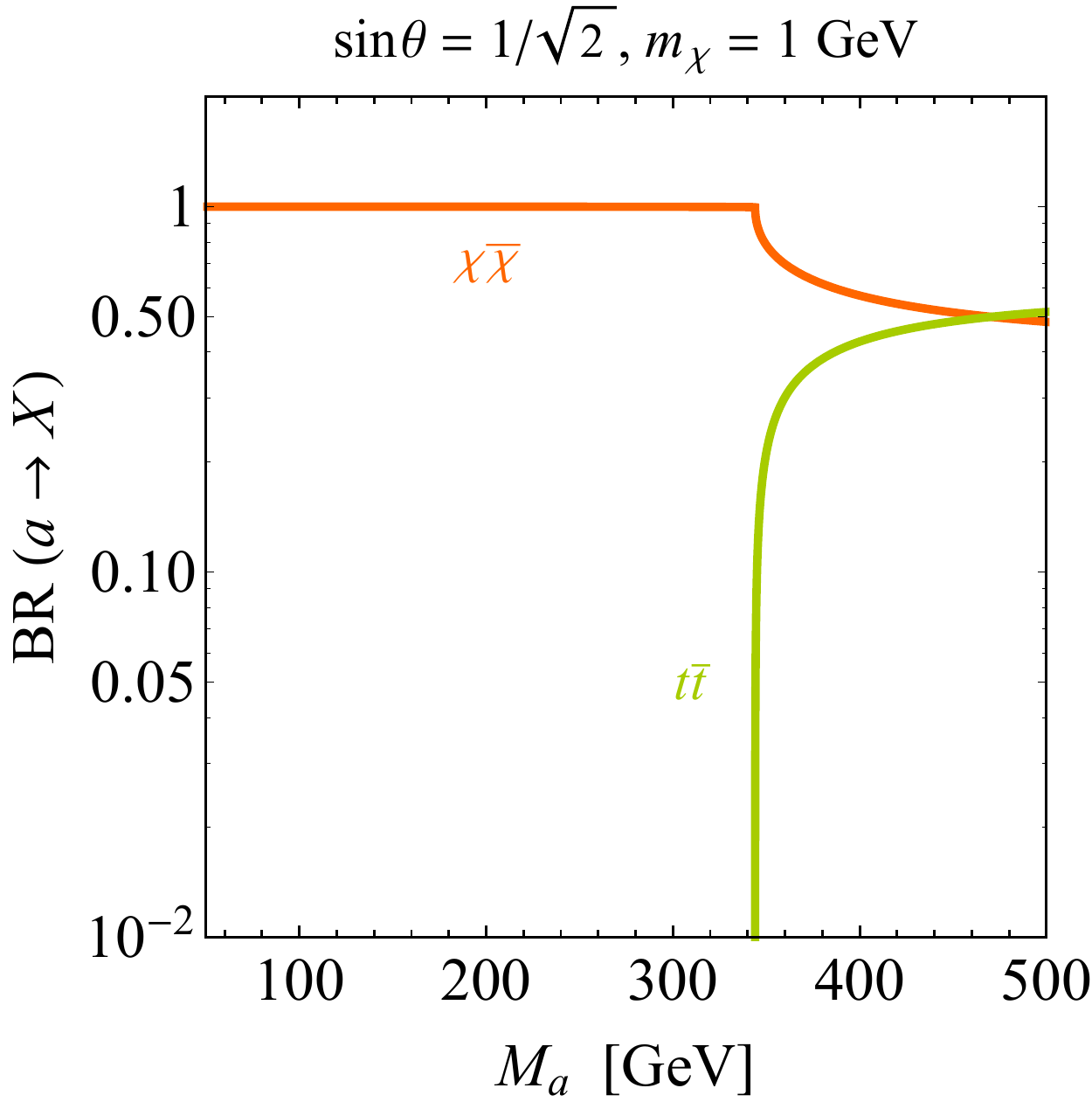} \qquad 
\includegraphics[width=0.425\textwidth]{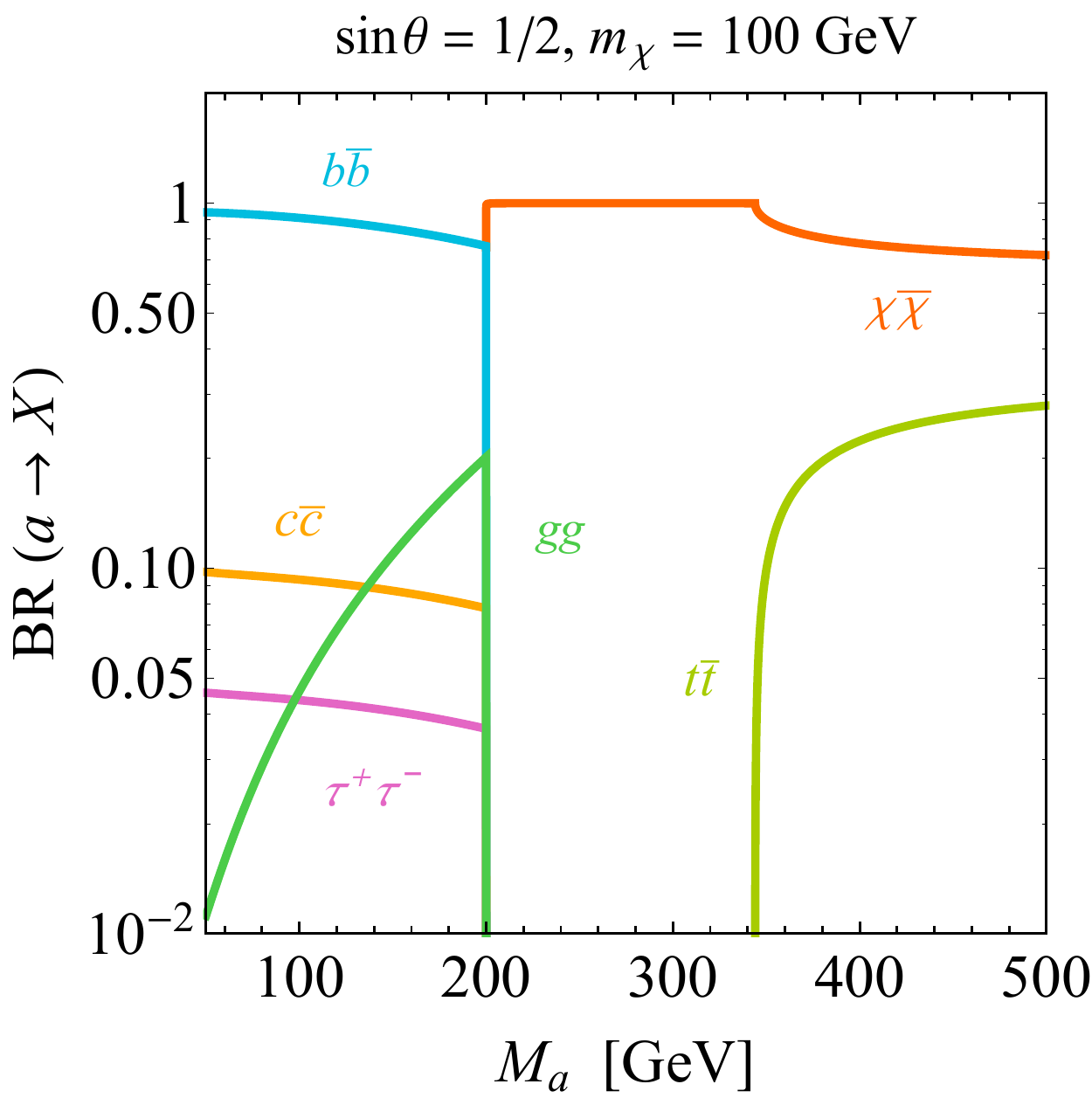}
\vspace{0mm}
\caption{Branching ratios of the lighter pseudoscalar $a$ as a function of its mass for two different choices of $\sin \theta$ and $m_\chi$ as indicated in the headline of the plots. The other relevant parameters have been set to $\tan \beta = 1$, $M_H = M_A = M_{H^\pm} = 750  \, {\rm GeV}$ and $y_\chi = 1$. Notice that for this specific $\tan \beta$ value the branching ratios of the pseudoscalar~$a$ do not depend on the choice of Yukawa sector. }
\label{fig:Bra}
\end{center}
\end{figure}

For small $\tan \beta$ and non-zero values of $\sin \theta$ the  couplings of $a$ to DM and top quarks dominate over all other couplings. As a result, the decay pattern of $a$ is in general very simple. This is illustrated in the panels of Figure~\ref{fig:Bra} for  two different  choices of parameter sets. The left panel shows the branching ratio of $a$ for a very light DM particle with~$m_\chi = 1 \, {\rm GeV}$.  One observes that below the $t \bar t$ threshold one has ${\rm BR} \left (a \to \chi \bar \chi \right) = 100 \%$ while for $M_a > 2 m_t$ both decays to DM and top-quarks pairs are relevant. In fact, sufficiently far above the $t \bar t$ threshold one obtains   ${\rm BR} \left (a \to \chi \bar \chi \right)/{\rm BR} \left (a \to t \bar t \right) \simeq 0.7 \hspace{0.5mm} y_\chi^2 \hspace{0.25mm} \tan^2 \beta/\tan^2 \theta$ independent of the specific realisation of the Yukawa sector. In the right panel we present our results for a DM state of $m_\chi = 100 \, {\rm GeV}$. In this case we see that below the $\chi \bar \chi$ threshold the pseudoscalar~$a$ decays  dominantly into bottom-quark pairs but that also the branching ratios to taus and gluons exceed the percent level. Compared to the left plot one also observes that in the right plot the ratio ${\rm BR} \left (a \to \chi \bar \chi \right)/{\rm BR} \left (a \to t \bar t \right)$ is significantly larger for $M_a > 2 m_t$ due to the different choice of $\sin \theta$. 

\subsection[Lighter scalar $h$]{Lighter scalar $\bm{h}$}
\label{sec:lighterscalar}

For sufficiently heavy pseudoscalars $a$ the decay pattern of $h$ resembles  that of the  SM Higgs boson in the alignment/decoupling limit. For $M_a < M_h/2$ on the other hand decays to two on-shell $a$ mediators are possible. The corresponding  partial decay width reads 
\beq \label{eq:Gammah1h4h4}
\Gamma \left ( h \to a a \right ) = \frac{1}{32 \pi} \, g_{haa}^2 \, M_h  \hspace{0.25mm}  \beta_{a/h} \,,
\eeq
with 
\beq \label{eq:ghaa}
\begin{split}
g_{haa} & = \frac{1}{M_h v}  \, \Big [  \left ( M_h^2  - 2 M_H^2  + 4 M_{H^\pm}^2 - 2 M_a^2 -  2 \lambda_3 \hspace{0.25mm} v^2 \right  ) \sin^2 \theta \\[2mm]
& \hspace{1.5cm} - 2 \left ( \lambda_{P1} \cos^2 \beta + \lambda_{P2} \sin^2 \beta \right ) v^2  \cos^2 \theta \hspace{0.5mm} \Big ] \,.
\end{split}
\eeq
Notice that the $haa$ coupling contains terms proportional to both $\sin^2 \theta$ and $\cos^2 \theta$. These contributions result from the trilinear and quartic couplings in the scalar potential (\ref{eq:VP}), respectively. In our THDM plus pseudoscalar extension, $h \to aa$ decays are even possible in the limit $\theta \to 0$, which is not the case in the simplified model considered in~\cite{Ipek:2014gua,No:2015xqa,Goncalves:2016iyg}.

\begin{figure}[t!]
\begin{center}
\includegraphics[width=0.425\textwidth]{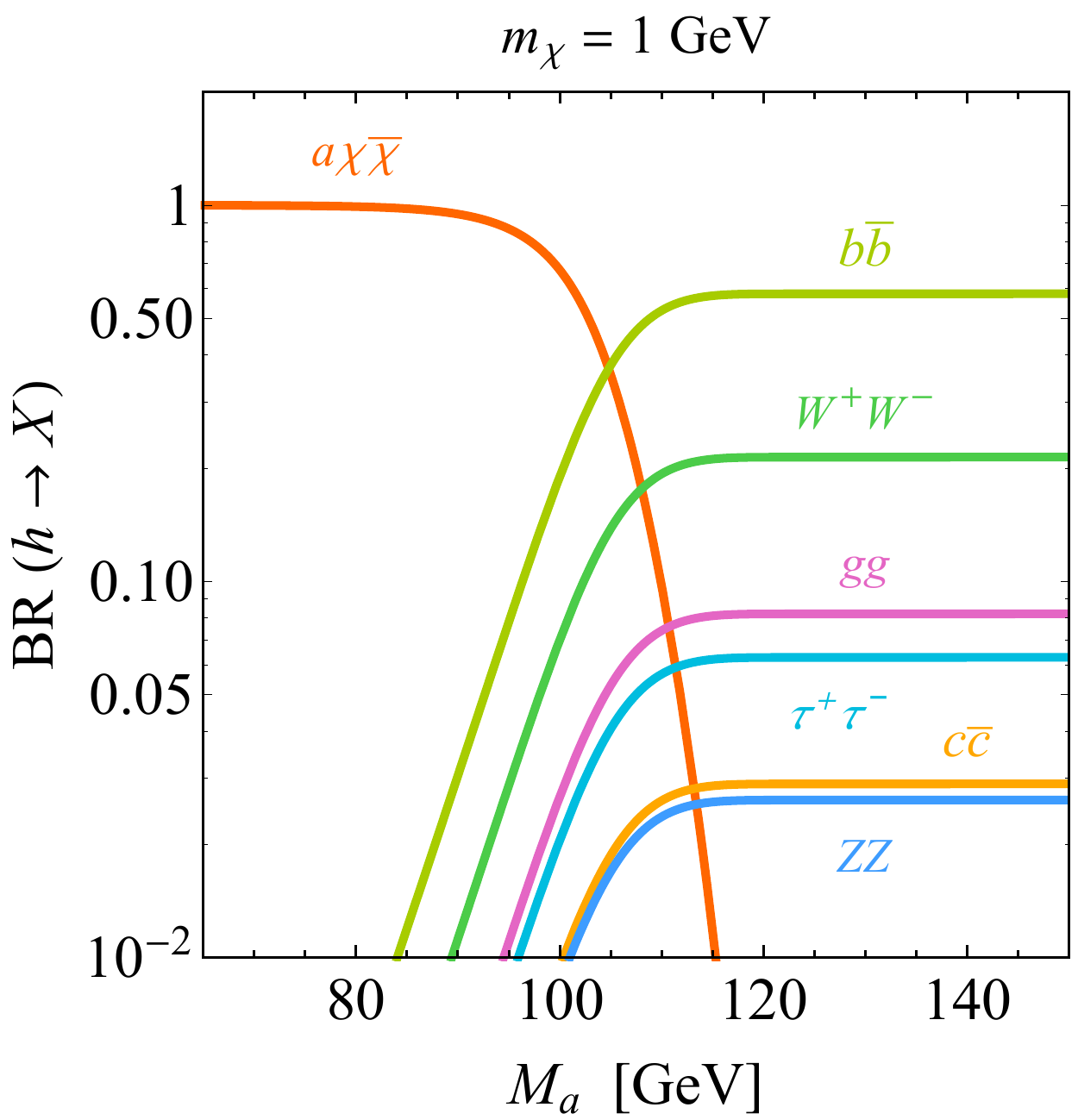} \qquad 
\includegraphics[width=0.425\textwidth]{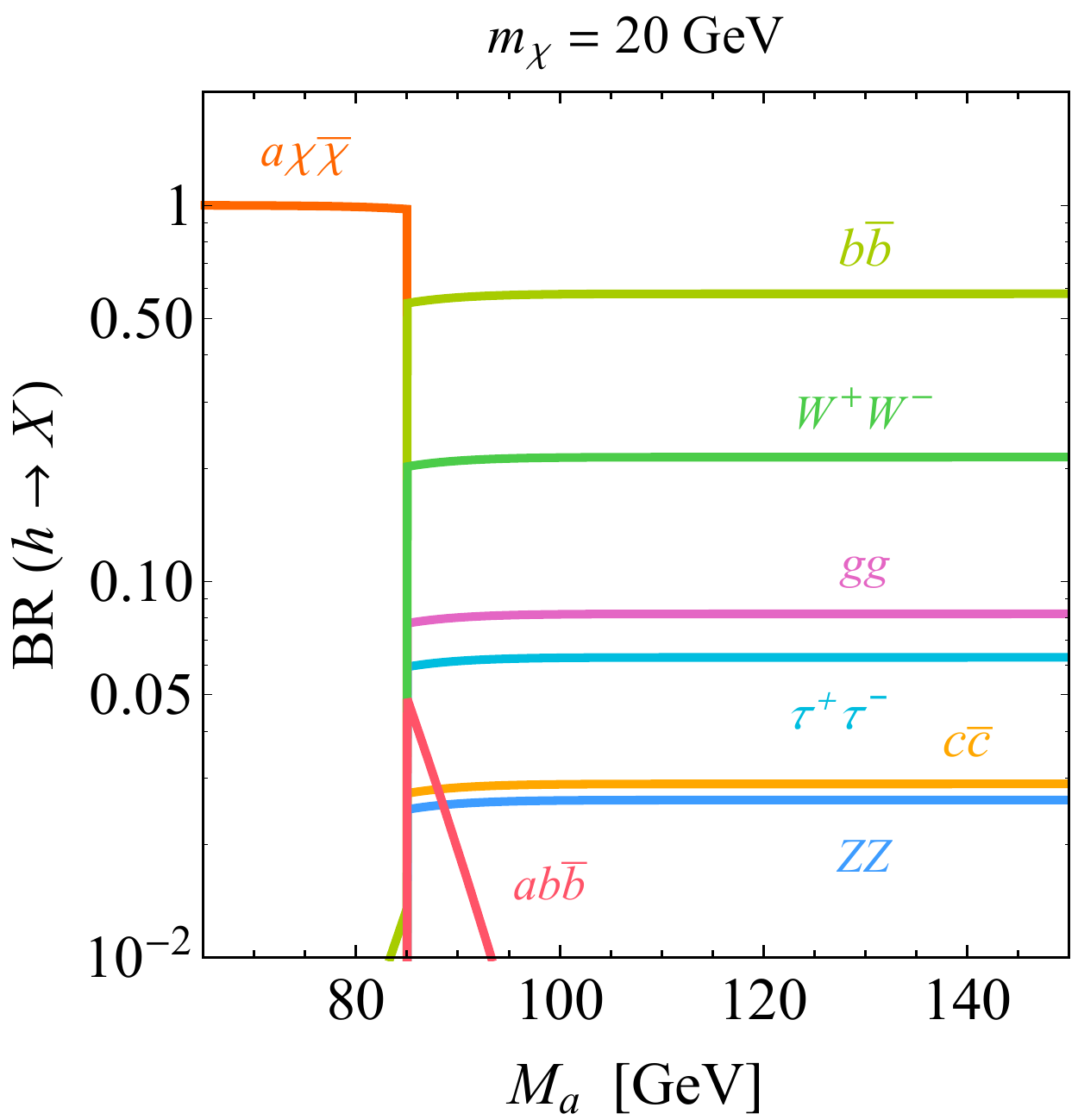}
\vspace{0mm}
\caption{Branching ratios of the lighter  scalar $h$ as a function of the pseudoscalar mass $M_a$ for two different choices of $m_\chi$ as indicated in the headline of the plots. The other relevant parameters have been set to  $\tan \beta  = 1$,  $M_H = M_A = M_{H^\pm} = 750  \, {\rm GeV}$, $\sin \theta = 1/\sqrt{2}$, $\lambda_3 = \lambda_{P1} = \lambda_{P2} = 0$ and~$y_\chi = 1$.}
\label{fig:Brh}
\end{center}
\end{figure}

Since the total decay width of the SM Higgs is only about $4 \, {\rm MeV}$, three-body decays of~$h$ into final states with a single $a$ can also be relevant in the mass range $M_h/2 < M_a \lesssim M_h$. Phenomenologically the most important three-body decay is the one where $a$ is accompanied by a  pair  of DM particles but decays to an $a$ and SM fermions are also possible. The corresponding partial decay widths are given by 
\beq \label{eq:Gammah1h4DMDMbb}
\begin{split}
\Gamma \left ( h \to a \chi \bar \chi \right ) & = \frac{y_\chi^2}{32 \pi^3} \, g_{haa}^2 \, M_h  \hspace{0.25mm}   \beta_{\chi/a} \hspace{0.5mm} g(\tau_{a/h}) \hspace{0.25mm} \cos^2 \theta\,, \\[2mm]
\Gamma \left ( h \to a f \bar f \right ) & = \frac{N_c^f \big ( \xi_f^{\rm M} \big)^2}{32 \pi^3} \frac{m_f^2}{v^2} \, g_{haa}^2 \, M_h  \hspace{0.25mm} \beta_{f/a} \hspace{0.5mm} g(\tau_{a/h}) \hspace{0.25mm} \sin^2 \theta \,, 
\end{split}
\eeq
with  \cite{Djouadi:1995gv}
\beq \label{eq:g3}
\begin{split}
g ( \tau  ) & = \frac{1}{8} \left ( \tau - 4  \right ) \left [ 4 -  \ln \left ( \frac{\tau}{4} \right )  \right ] \\[2mm] & \phantom{xx} - \frac{5 \tau - 4} {4 \sqrt{\tau-1}} \left [ \arctan \left ( \frac{\tau-2}{2 \sqrt{\tau-1}} \right ) - \arctan \left ( \frac{1}{\sqrt{\tau-1}} \right )\right ] \, . 
\end{split}
\eeq

In Figure~\ref{fig:Brh} we show the branching ratios of the SM Higgs $h$ for two different values of the~DM mass. We observe that for a light pseudoscalar mediator $a$  one has in both cases ${\rm BR} \left ( h \to a \chi \bar \chi \right ) = 100 \%$. In fact, the total decay width of the lighter scalar $h$ exceeds $3 \, {\rm GeV}$ for masses $M_a \lesssim 70 \, {\rm GeV}$. Such large values of  $\Gamma_h$ are in conflict with the model-independent upper limits on the total decay width of the Higgs as measured by both ATLAS and CMS in LHC~Run~I~\cite{Aad:2014aba,Khachatryan:2014jba}. Notice that since the pseudoscalar $a$ decays with 100\% to DM pairs for the considered values of $m_\chi$ one has ${\rm BR} \left ( h \to a \chi \bar \chi \right ) = {\rm BR} \left ( h \to 2 \chi 2 \bar \chi \right )$. This implies that for light DM the simplified model presented in Section~\ref{sec:THDMP} is subject to the constraints arising from invisible decays of the Higgs boson \cite{Aad:2015pla,Khachatryan:2016whc}. We will analyse the resulting restrictions on the parameter space in~Section~\ref{sec:wetdream}. The right panel  finally illustrates that in cases where~$m_\chi$ is close to a quarter of the SM Higgs mass also decays such as $h \to a b \bar b$ with $a \to \chi \bar \chi$ can have branching ratios of a few percent (or more) for a narrow range of~$M_a$ values.  Notice that for the choice $\tan \beta = 1$ used in the figure the result for ${\rm BR} \left ( h \to a b \bar b \right )$ does not depend on the particular Yukawa assignment. 

\subsection[Heavier scalar $H$]{Heavier scalar $\bm{H}$}
\label{sec:heavierscalar}

In the alignment/decoupling limit of the pseudoscalar extensions of the THDM model the heavier scalar $H$ does not couple to $W^+ W^-$ and $Z Z$ pairs. In addition the $Hhh$  vertex vanishes. Under the assumption that $M_H > M_a$ and taking $A$ to be sufficiently heavy, the scalar $H$ can hence decay only to SM fermions or the $aa$ and $aZ$ final state at tree level. The corresponding partial decay widths are 
\beq \label{eq:GammaHX}
\begin{split}
\Gamma \left ( H \to f \bar f \right ) & =  \frac{N_c^f \big ( \xi_f^{\rm M} \big)^2}{8\pi}  \hspace{0.1mm}  \frac{m_f^2}{v^2}  \hspace{0.25mm}  M_H  \hspace{0.25mm}  \beta_{f/H}^3  \,, \\[2mm]
\Gamma \left ( H \to a a \right ) & =  \frac{1}{32 \pi}  \, g_{Haa}^2 \, M_H  \hspace{0.25mm}  \beta_{a/H} \,, \\[2mm]
\Gamma \left ( H \to a Z \right ) & =   \frac{1}{16 \pi} \frac{\lambda^{3/2} (M_H, M_a, M_Z)}{M_H^3 v^2} \, \sin^2 \theta \,,
\end{split}
\eeq
with 
\beq \label{eq:gHaa}
\begin{split}
g_{Haa}  & = \frac{1}{M_H \hspace{0.125mm} v} \, \Big [  \cot \left ( 2 \beta \right) \left (  2 M_h^2 - 4 M_H^2 + 4 M_{H^\pm}^2  - 2 \lambda_3 v^2 \right )  \sin^2 \theta  \\  & \hspace{1.6cm} + \sin \left ( 2 \beta \right ) \left (\lambda_{P1}-\lambda_{P2} \right ) v^2 \cos^2 \theta  \hspace{0.5mm} \Big  ] \,,
\end{split}
\eeq
denoting the $Haa$ coupling. We have furthermore introduced 
\beq \label{eq:lambda}
\lambda (m_1, m_2, m_3) = \left ( m_1^2 - m_2^2 - m_3^2 \right )^2 - 4 \hspace{0.25mm}  m_2^2 \hspace{0.5mm} m_3^2 \,,
\eeq
which characterises the two-body phase space for three massive particles. Notice that the appearance of $\lambda_{P1}$  and  $\lambda_{P2}$  in the partial decay width $\Gamma \left ( H \to a a \right )$ indicates again a qualitative difference between the scalar interactions considered  in~\cite{Ipek:2014gua,No:2015xqa,Goncalves:2016iyg} and the more general potential~(\ref{eq:VP}). At the one-loop level the heavier scalar $H$ can in addition decay  to gluons and other gauge bosons, but the associated branching ratios are very suppressed and  thus  have no impact on our numerical results. 

\begin{figure}[t!]
\begin{center}
\includegraphics[width=0.425\textwidth]{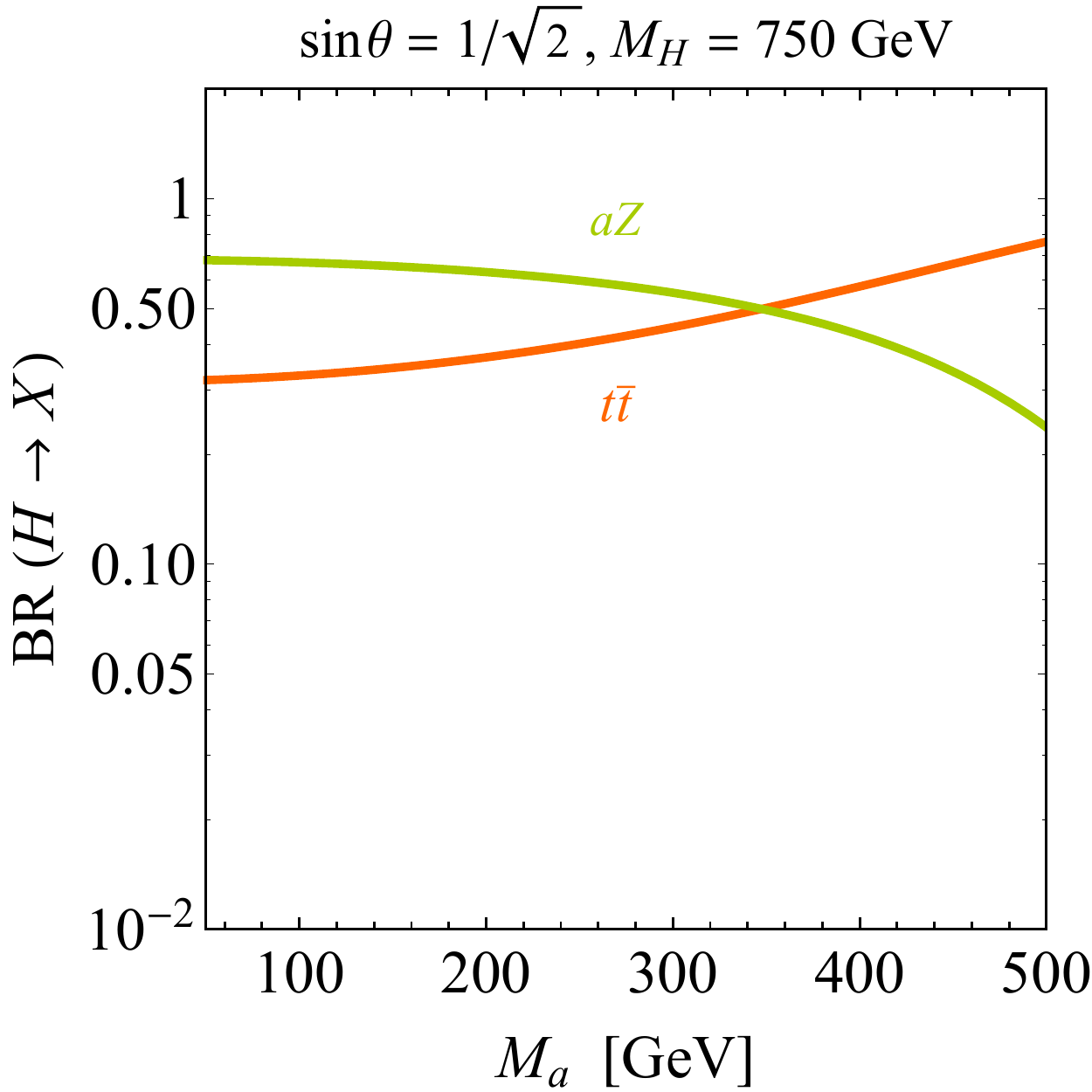} \qquad 
\includegraphics[width=0.425\textwidth]{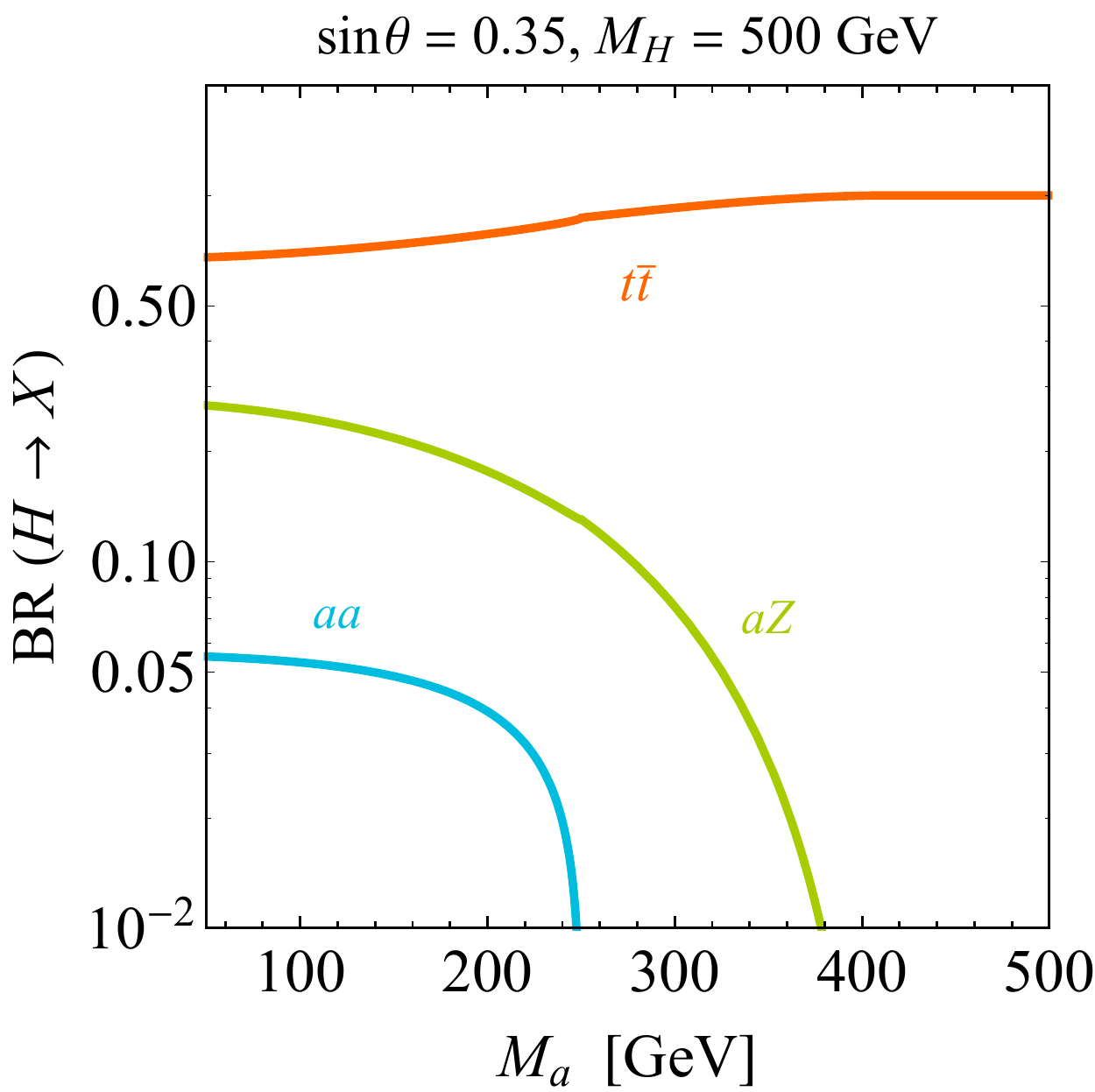}
\vspace{0mm}
\caption{Branching ratios of the heavier scalar $H$ as a function of  $M_a$ for two different choices of $\sin \theta$ and $M_{H}$ as indicated in the headline of the plots. The other used input parameters are $\tan \beta = 1$,  $M_A = M_{H^\pm} = 750 \, {\rm GeV}$, $\lambda_3 =  \lambda_{P2} = 0$ and $\lambda_{P1} = 1$. }
\label{fig:BrH}
\end{center}
\end{figure}

The dominant branching ratios of $H$ as a function of $M_a$ are displayed in Figure~\ref{fig:BrH} for two parameter sets. In the left panel the case of a scalar $H$ with $\sin \theta = 1/\sqrt{2}$ and $M_H = 750 \, {\rm GeV}$ is shown. One observes that for $M_a \lesssim 350 \, {\rm GeV}$ the decay mode $H \to aZ$ has the largest branching ratio,  while  for heavier $a$ the $H \to t \bar t$ channel represents the leading decay. Notice that for model realisations where the decay channel $H \to a Z$ dominates, interesting~mono-$Z$ signatures can be expected~\cite{No:2015xqa,Goncalves:2016iyg}. We will come back to this point in Section~\ref{sec:monoZ}. The decay pattern of $H$ is however strongly dependent on the mass of~$H$ since for $M_H < M_{H^\pm}$ the mixing angle $\theta$ is constrained to be small by EW precision measurements (see Section~\ref{sec:EWprecision}). This behaviour is easy to understand from  (\ref{eq:GammaHX})  which in the  limit of small $\sin \theta$, $\tan \beta = {\cal O} ( 1)$ and large~$M_H$ imply that $\Gamma \left ( H \to t \bar t \right ) \propto m_t^2/(M_H \tan^2 \beta)$, $\Gamma \left ( H \to a a \right ) \propto v^4/M_H^3 \left ( \lambda_{P1} - \lambda_{P2} \right )^2$  and $\Gamma \left ( H \to a Z \right ) \propto M_H \sin^2 \theta$. For $M_H > 2 m_t$  the decay mode~$H \to t \bar t$ can hence  dominate over the whole $M_a$ range of   interest. This feature is illustrated on the right-hand side of the figure for $\sin \theta = 0.35$ and $M_H = 500 \, {\rm GeV}$. One also sees from this panel that    the  branching ratio of $H \to aa$ can be relevant as it does not tend to zero in the $\sin \theta \to 0$ limit if the combination $ \lambda_{P1} - \lambda_{P2} $ of quartic couplings is non-zero. For $\tan \beta \gtrsim 2$ and $ \lambda_{P1} - \lambda_{P2} \gtrsim 1$,  ${\rm BR} \left ( H \to a a \right)$   can even be the largest  branching ratio for $M_a < M_H/2$. This happens because the terms proportional to $\sin^2 \theta$ and $\cos^2 \theta$ in (\ref{eq:gHaa}) both give a sizeable contribution to the $Haa$~coupling, while the $H t \bar t$ coupling is suppressed by $1/\tan^2 \beta$. 

\subsection[Heavier pseudoscalar $A$]{Heavier pseudoscalar $\bm{A}$}
\label{sec:heavierpseudoscalar}

For $M_A > M_a$  and assuming that decays to $H$ are kinematically inaccessible, the pseudoscalar $A$ can only decay to DM, SM fermions and the $ah$ final state at tree level. In the alignment/decoupling limit the corresponding partial decay widths take the form 
\beq \label{eq:GammaAX}
\begin{split}
\Gamma \left ( A \to \chi \bar \chi \right ) & = \frac{y_\chi^2}{8\pi} \hspace{0.25mm} M_A  \hspace{0.25mm}  \beta_{\chi/A} \sin^2 \theta \,,  \\[2mm]
\Gamma \left ( A \to f \bar f \right ) & =  \frac{N_c^f \big ( \xi_f^{\rm M} \big)^2}{8\pi}  \hspace{0.1mm}  \frac{m_f^2}{v^2}  \hspace{0.25mm}  M_A  \hspace{0.25mm}  \beta_{f/A}   \cos^2 \theta \,, \\[2mm]
\Gamma \left ( A \to a h  \right ) & =   \frac{1}{16 \pi} \frac{\lambda^{1/2} (M_A, M_a, M_h)}{M_A} \, g_{Aah}^2  \,,
\end{split}
\eeq
with 
\beq \label{eq:gAah}
\begin{split}
g_{Aah}  & = \frac{1}{M_A v} \, \Big [ \hspace{0.5mm} M_h^2  - 2 M_H^2  - M_A^2 + 4 M_{H^\pm}^2 -  M_a^2 - 2 \lambda_3 v^2  \\[2mm] & \hspace{1.6cm} + 2 \left (  \lambda_{P1} \cos^2 \beta + \lambda_{P2} \sin^2 \beta  \right ) v^2 \hspace{0.5mm} \Big ] \sin \theta \cos \theta \,,
\end{split}
\eeq
denoting the $Aah$ coupling, and the analytic expression for the two-body phase-space  function $\lambda (m_1, m_2, m_3)$ can be found in (\ref{eq:lambda}). Like in the case of~$H$, loop-induced decays of the heavier pseudoscalar~$A$ can be neglected for all practical purposes. 

\begin{figure}[t!]
\begin{center}
\includegraphics[width=0.425\textwidth]{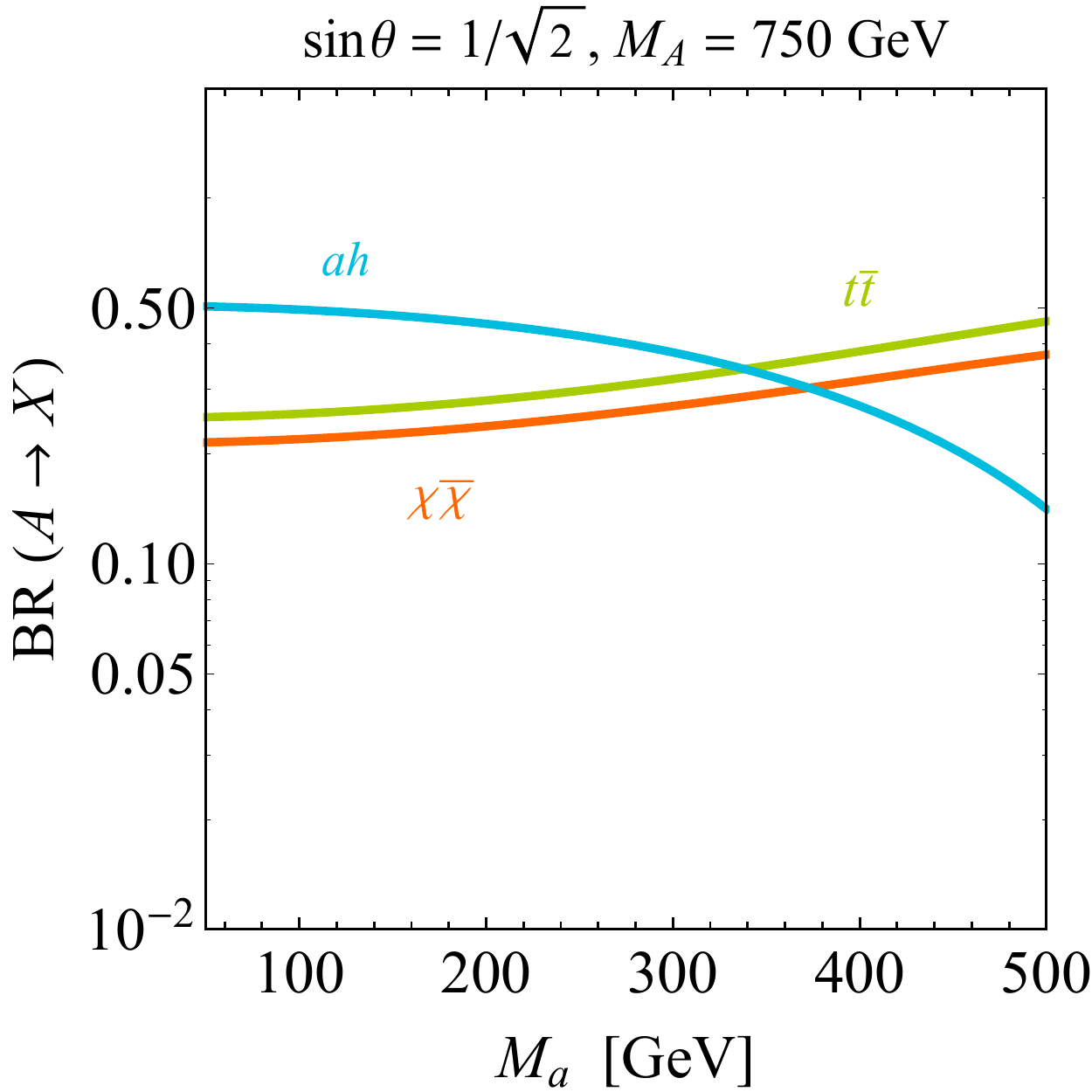} \qquad 
\includegraphics[width=0.425\textwidth]{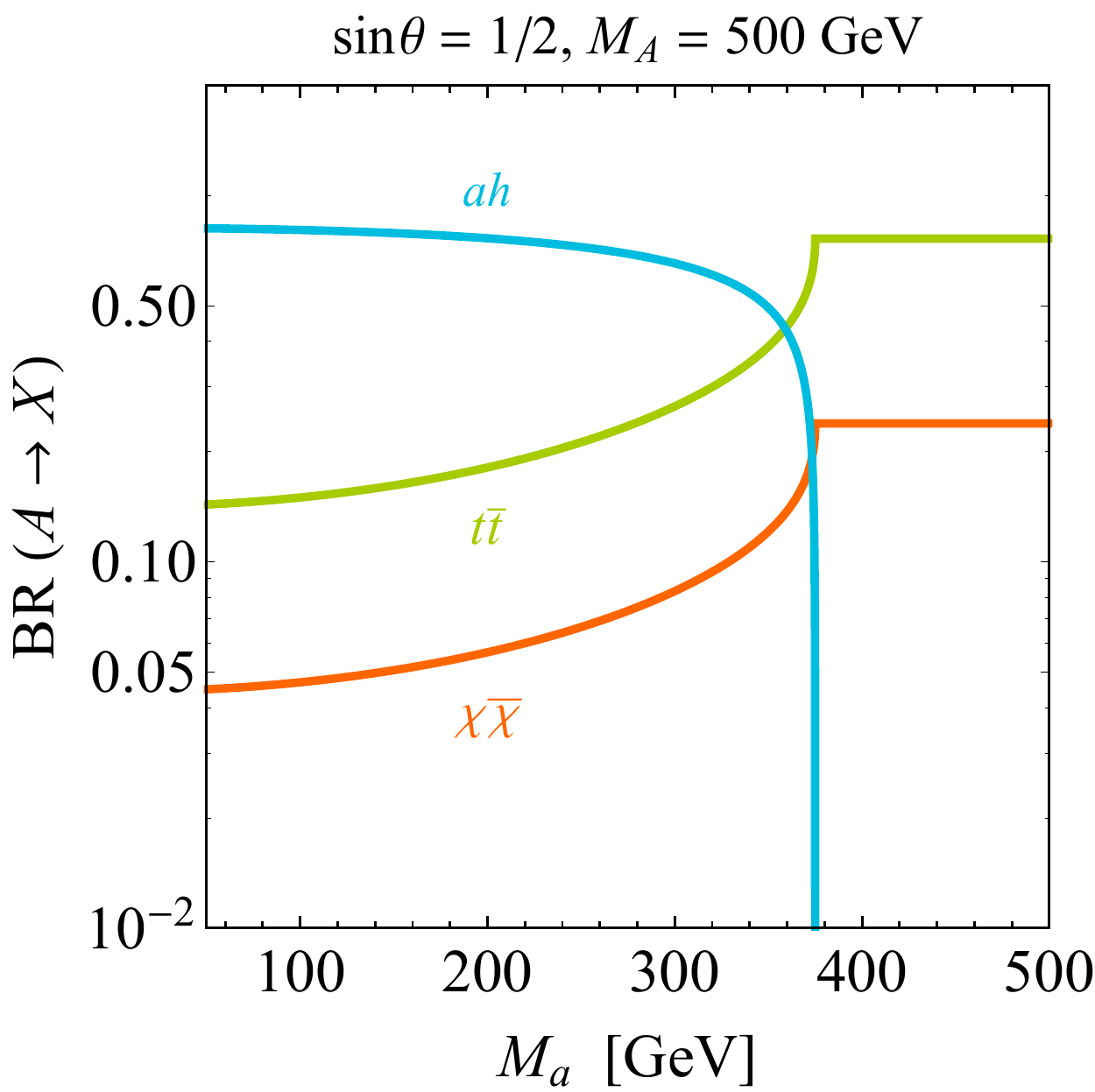}
\vspace{0mm}
\caption{Branching ratios of the heavier pseudoscalar $A$ as a function of  $M_a$ for two different choices of $M_A$ and $\sin \theta$ as indicated in the headline of the plots. The other parameter choices are  $\tan \beta = 1$,  $M_H = M_{H^\pm} = 750 \, {\rm GeV}$, $\lambda_3 = \lambda_{P1} = \lambda_{P2} = 0$, $y_\chi = 1$ and $m_\chi= 1 \, {\rm GeV}$.}
\label{fig:BrA}
\end{center}
\end{figure}

In Figure~\ref{fig:BrA} we present our results for the  branching ratios of the pseudoscalar $A$ as a function of $M_a$ for two different parameter choices. The left panel illustrates the case $M_A = 750 \, {\rm GeV}$ and one sees that for such an $A$ the  branching ratios are all above 10\% and  the hierarchy ${\rm BR} \left ( A \to ah \right ) > {\rm BR} \left ( A \to t \bar t \right ) > {\rm BR} \left ( A \to \chi \bar \chi \right )$ is observed for $M_a \lesssim 200 \, {\rm GeV}$. As shown on the  right-hand side of the figure, this hierarchy not only remains intact   but is even more pronounced for a moderately heavy $A$ until the threshold $M_a = M_A - M_h$ is reached. For larger $M_a$ values only decays to $\chi \bar \chi$ and $t \bar t$ final states matter and the ratio of their  branching ratios is approximately given by ${\rm BR} \left (A \to \chi \bar \chi \right)/{\rm BR} \left (A \to t \bar t \right) \simeq 0.9 \hspace{0.5mm} y_\chi^2 \hspace{0.25mm} \tan^2 \beta \hspace{0.25mm} \tan^2 \theta$ irrespective of the particular Yukawa assignment. Notice that a sizeable $A \to ah$ branching ratio is a generic prediction in the THDM plus pseudoscalar extensions with small $\tan \beta$, since the charged Higgs has to be quite heavy in this case in order to avoid the bounds from $ B \to X_s \gamma$ and/or $B_s$-meson mixing.   Since $a  \to \chi \bar \chi$ is typically the dominant decay mode of the lighter  pseudoscalar $a$, appreciable mono-Higgs signals are hence a firm prediction in a certain region of parameter space of our simplified model. This point will be further explained in Section~\ref{sec:monohiggs}. 

\subsection[Charged scalar $H^\pm$]{Charged scalar $\bm{H^\pm}$}
\label{sec:chargedscalar}

Since in the alignment/decoupling limit the $H^+ h W^+$ vertex vanishes, the  partial decay widths of the charged scalar $H^+$ that are relevant  in the small $\tan \beta$ regime  read 
\beq \label{eq:GammaHpX}
\begin{split}
\Gamma \left ( H^+ \to t \bar b \right ) & = \frac{N_c^t   \hspace{0.75mm}  |V_{tb}|^2 \big ( \xi_t^{\rm M} \big)^2}{8\pi} \hspace{0.25mm} \frac{m_t^2}{v^2}  \hspace{0.5mm} M_{H^\pm} \left ( 1 - \frac{m_t^2}{ M_{H^\pm}^2} \right )^2 \,, \\[2mm]
\Gamma \left ( H^+ \to H W^+ \right ) & = \frac{1}{16\pi} \hspace{0.25mm} \frac{\lambda^{3/2} (M_{H^\pm}, M_H, M_W)}{M_{{H^\pm}}^3 v^2}   \,,  \\[2mm]
\Gamma \left ( H^+ \to A W^+ \right ) & = \frac{1}{16\pi} \hspace{0.25mm} \frac{\lambda^{3/2} (M_{H^\pm}, M_A, M_W)}{M_{{H^\pm}}^3 v^2} \, \cos^2 \theta  \,,  \\[2mm]
\Gamma \left ( H^+ \to a W^+ \right ) & = \frac{1}{16\pi} \hspace{0.25mm} \frac{\lambda^{3/2} (M_{H^\pm}, M_a, M_W)}{M_{{H^\pm}}^3 v^2} \, \sin^2 \theta  \,, 
\end{split}
\eeq
where in the case of $H^+ \to t \bar b$ we have  neglected terms  of ${\cal O} (m_b^2/M_{H^\pm}^2)$ in the expression for the partial decay width. Notice that in THDMs of type II and III also the decay $H^+ \to \tau^+ \nu_\tau$ can be important if $\tan \beta \gg 1$. The result for $\Gamma \left (H^+ \to \tau^+ \nu_\tau \right )$ can be obtained from the expression given above for $\Gamma \left ( H^+ \to t \bar b \right )$ by obvious replacements.

\begin{figure}[t!]
\begin{center}
\includegraphics[width=0.425\textwidth]{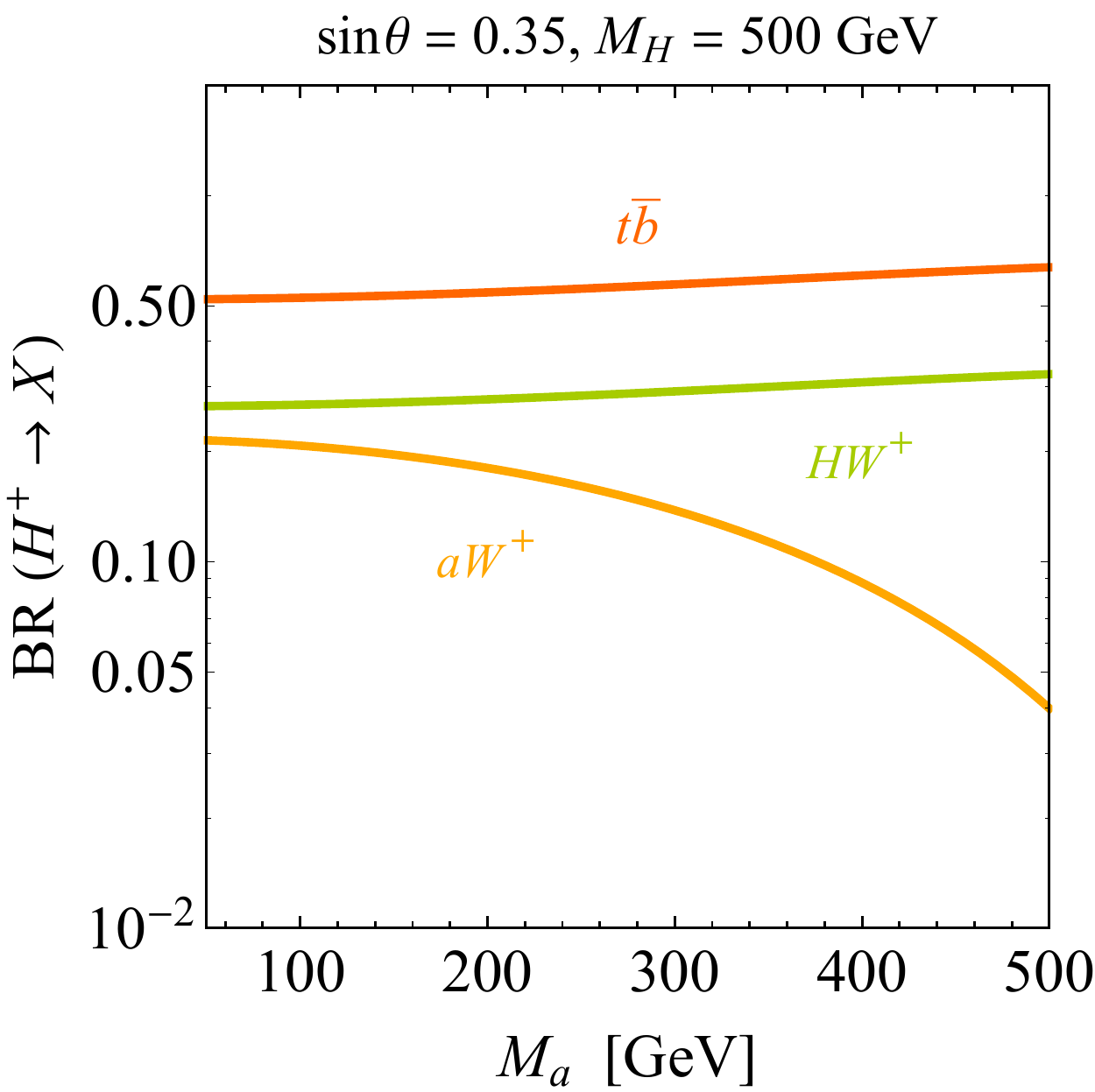} \qquad 
\includegraphics[width=0.425\textwidth]{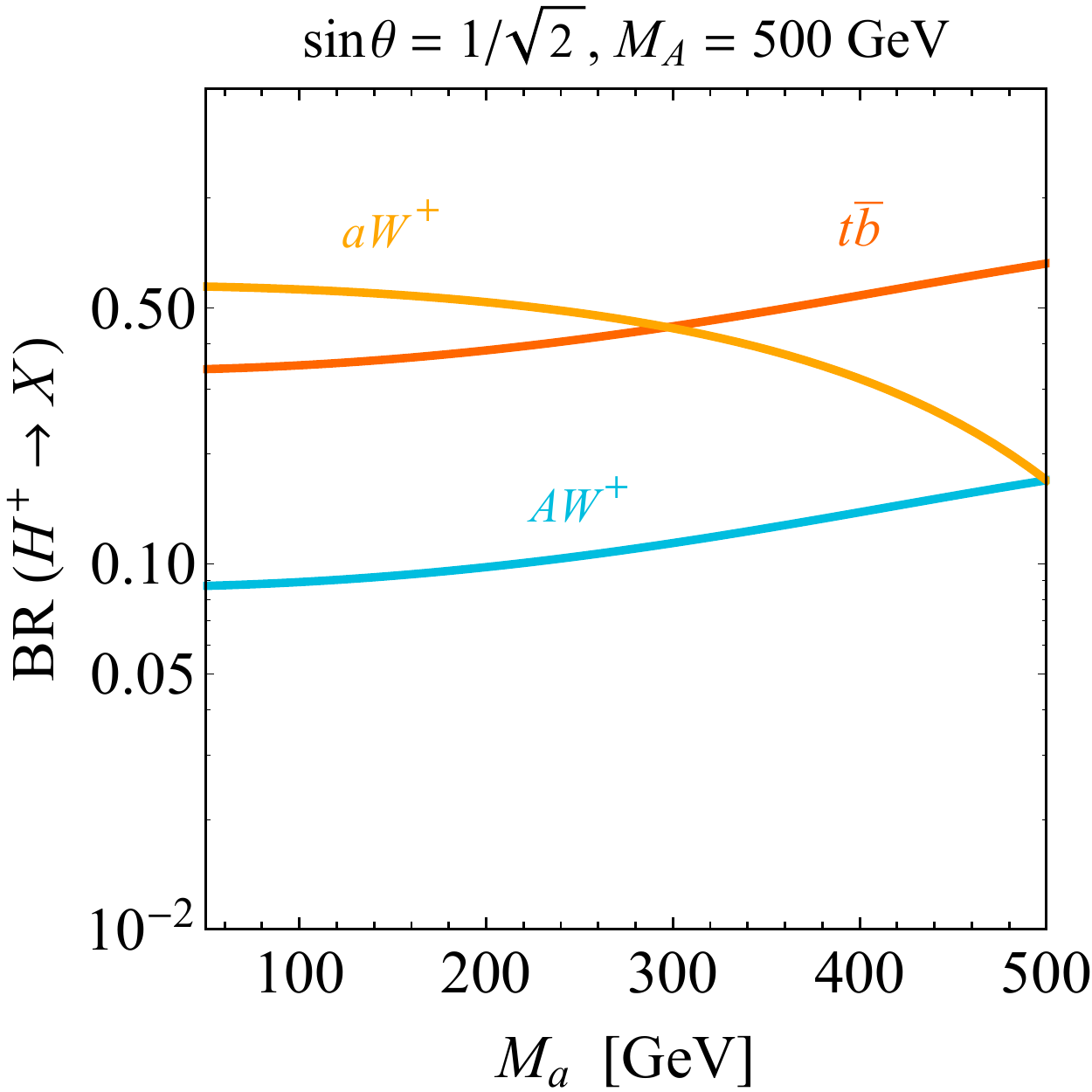}
\vspace{0mm}
\caption{Branching ratios of the charged scalar $H^+$ as a function of  $M_a$ for two different sets of input parameters as indicated in the headline of the plots. In the left (right) panel in addition $\tan \beta = 1$  and $M_A = M_{H^\pm} = 750 \, {\rm GeV}$ ($M_H = M_{H^\pm} = 750 \, {\rm GeV}$) is used.}
\label{fig:BrHc}
\end{center}
\end{figure}

The main branching ratios of the charged Higgs $H^+$ are displayed in Figure~\ref{fig:BrHc}. On the left-hand side of the figure the case of $\sin \theta = 0.35$ and $M_H = 500 \, {\rm GeV}$  is displayed and one observes that ${\rm BR} \left (H^+ \to t \bar b \right ) > {\rm BR} \left (H^+ \to HW^+ \right ) > {\rm BR} \left (H^+ \to aW^+ \right )$ for the shown values of $M_a$. Notice that for scenarios with $M_H < M_{H^\pm}$ the hierarchy ${\rm BR} \left (H^+ \to HW^+ \right ) > {\rm BR} \left (H^+ \to aW^+ \right )$ is a rather model-independent prediction since in such cases EW precision measurements require $\sin \theta$ to be small and $\Gamma \left ( H^+ \to a W^+ \right )/ \Gamma \left ( H^+ \to H W^+ \right ) \propto \sin^2 \theta$. The same is not true for the hierarchy between ${\rm BR} \left (H^+ \to t \bar b \right )$ and ${\rm BR} \left (H^+ \to HW^+ \right )$ which depends sensitively on the choice of $\tan \beta$ since $\Gamma \left ( H^+ \to t \bar b  \right )/ \Gamma \left ( H^+ \to H W^+ \right ) \propto 1/\tan^2 \beta$. It follows that for values of $\tan \beta > 1$ the $H^+ \to H W^+$ channel can also be the dominant decay mode. In model realisations with $M_A < M_{H^\pm}$ there are no constraints from~$\Delta \rho$ on $\sin \theta$ and in turn the $H^+ \to aW^+$ branching ratio can dominate for sufficiently large mixing in the pseudoscalar sector. This feature is illustrated by the right panel in the figure using $\sin \theta = 1/\sqrt{2}$ and $M_A= 500 \, {\rm GeV}$. For this choice of input parameters we find that ${\rm BR} \left (H^+ \to aW^+ \right ) > {\rm BR} \left (H^+ \to t \bar b \right )$ for masses $M_a \lesssim 300 \, {\rm GeV}$. Since the  pseudoscalar~$a$ predominantly decays via $a \to \chi \bar \chi$ it follows that THDM plus pseudoscalar extensions with $M_A < M_{H^\pm}$ can lead to a resonant mono-$W$ signal. We will discuss the LHC prospects for the detection of such a $E_{T,\rm miss}$ signature in Section~\ref{sec:monoW}. 
 
\section{Anatomy of mono-$\bm{X}$ signatures}
\label{sec:METsignals}

In this section we will discuss the most important features of the mono-$X$ phenomenology of the pseudoscalar extensions of the THDM. We examine the mono-jet, the $t \bar t + E_{T, \rm miss}$, the mono-$Z$ and the mono-Higgs signature. The $b \bar b + E_{T, \rm miss}$ and mono-$W$  channel are also briefly considered. Our numerical analysis of the mono-$X$ signals is postponed to Section~\ref{sec:numerics}.

\subsection{Mono-jet channel}
\label{sec:monojet}

\begin{figure}[!t]
\begin{center}
\includegraphics[height=0.225\textwidth]{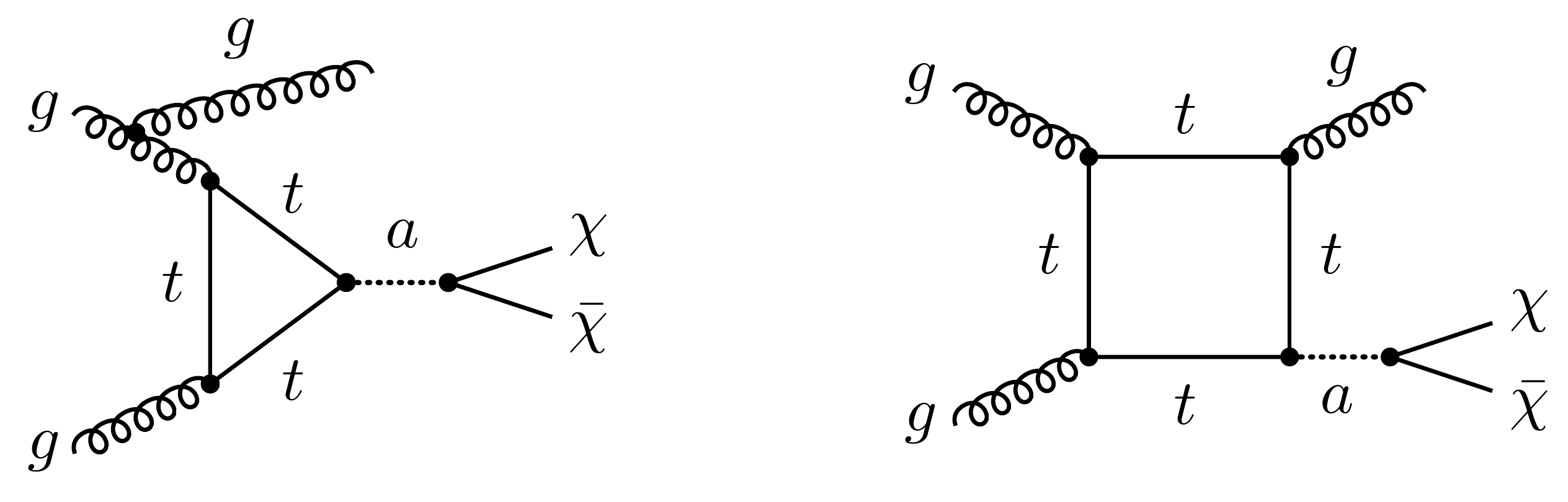}  
\vspace{0mm}
\caption{\label{fig:jDMDM} Examples of  diagrams that give rise to  a $j +E_{T,\rm miss}$ signature through the exchange of a lighter pseudoscalar $a$. Graphs involving a heavier pseudoscalar $A$ also contribute to the signal in the pseudoscalar extensions of the THDM but are not shown explicitly. }
\end{center}
\end{figure}

A first possibility to search for pseudoscalar interactions of the form~(\ref{eq:fermioninteractions}) consists in looking for a mono-jet signal, where the mediators that pair produce DM are radiated from heavy-quark loops~\cite{Haisch:2015ioa,Haisch:2012kf,Fox:2012ru,Haisch:2013ata,Haisch:2013fla,Buckley:2014fba,Harris:2014hga,Mattelaer:2015haa,Arina:2016cqj}. Representative examples of the possible one-loop Feynman diagrams are shown in Figure~\ref{fig:jDMDM}. 

For $m_a > 2 m_\chi$ and $M_A\gg M_a$ only graphs involving the exchange of the light pseudoscalar $a$ will contribute to the $j + E_{T, \rm miss}$ signal. As a result the normalised kinematic distributions of the mono-jet signal in the pseudoscalar extensions of the THDM are identical to those of the DMF pseudoscalar model.  Working in the NWA and assuming that $\tan \beta$ is small, the ratio of the fiducial cross sections in the two models is thus approximately given by the simple expression 
\beq \label{eq:monojetratio}
\frac{\sigma \left (pp \to j + E_{T, \rm miss} \right )}{\sigma \left (pp \to j + E_{T, \rm miss} \right )_{\rm DMF}} \simeq \left ( \frac{y_\chi \sin \theta}{g_\chi \hspace{0.25mm} g_q \tan \beta} \right )^2 \,.
\eeq
Here $g_\chi$ ($g_q$) denotes the DM-mediator (universal quark-mediator) coupling in the corresponding DMF spin-0 simplified model. Notice that the above relation is largely independent of the choice of Yukawa sector as long as $\tan \beta = {\cal O} (1)$ since bottom-quark loops have only an effect of a few percent on the $j + E_{T, \rm miss}$ distributions  (see for instance \cite{Bishara:2016jga} for a related discussion in the context of Higgs physics). Using the approximation (\ref{eq:monojetratio})  it is straightforward to recast existing mono-jet results on the DMF pseudoscalar model such as those given in \cite{CMS:2016pod} into the THDM plus pseudoscalar model space. The numerical results presented in the next section however do not employ any approximation since they are based on   a calculation of the $j + E_{T, \rm miss}$  cross sections including both top-quark and bottom-quark loops as well as the exchange of both $a$ and $A$ mediators. 

\subsection[$t \bar t/b \bar b + E_{T, \rm miss}$ channels]{$\bm{t \bar t/b \bar b + E_{T, \rm miss}}$ channels}
\label{sec:ttbbassociated}

\begin{figure}[!t]
\begin{center}
\includegraphics[height=0.225\textwidth]{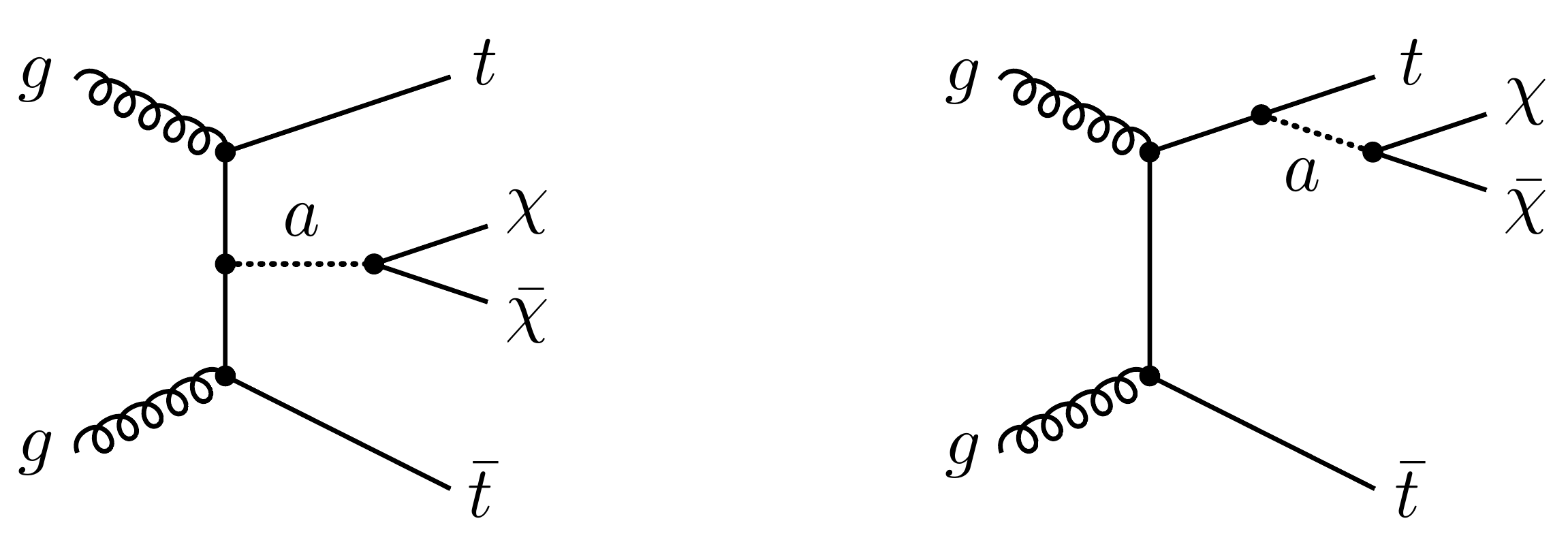}  
\vspace{0mm}
\caption{\label{fig:ttDMDM} Two possible diagrams that give rise to  a $t \bar t +E_{T,\rm miss}$ signal. Graphs with both an exchange of an $a$ and $A$ contribute in the THDM plus pseudoscalar extensions but only the former are displayed. }
\end{center}
\end{figure}

A second channel that is known to be a sensitive probe of top-philic pseudoscalars with large invisible decay widths is associated production of DM  and $t \bar t$ pairs~\cite{Haisch:2015ioa,Buckley:2014fba,Arina:2016cqj,Lin:2013sca,Artoni:2013zba,Backovic:2015soa,Haisch:2016gry}. Figure~\ref{fig:ttDMDM} displays examples of tree-level diagrams that give rise to a $t \bar t + E_{T, \rm miss}$ signature in the pseudoscalar extensions of the THDM model. 

In the case that $A$ is again much heavier than $a$, the signal strength for $t \bar t + E_{T, \rm miss}$ in our simplified model can be obtained from the prediction in the DMF pseudoscalar scenario from a rescaling relation analogous to the one shown in (\ref{eq:monojetratio}). Using such a simple recasting procedure we find that the most recent  ATLAS~\cite{ATLAS:2016ljb} and CMS searches for $t \bar t+ E_{T, \rm miss}$~\cite{CMS:2016mxc}  that are based on $13.2 \, {\rm fb}^{-1}$ and $2.2 \, {\rm fb}^{-1}$ of 13~TeV LHC data, respectively, only allow to set very weak bounds on $\tan \beta$.  For instance for $M_a = 100 \, {\rm GeV}$, $y_\chi =1$ and $m_\chi = 1 \, {\rm GeV}$  a lower limit of $\tan \beta \gtrsim 0.2$ is obtained. The  $t \bar t+ E_{T, \rm miss}$ constraints on the parameter space of the pseudoscalar extensions of the THDM are however expected to improve notably at forthcoming LHC runs. The numerical results that will be presented in Section~\ref{sec:wetdream} are based on the search strategy developed recently in~\cite{Haisch:2016gry} which employs a shape fit to the difference in pseudorapidity of the two charged leptons in the di-leptonic channel of $t \bar t+ E_{T, \rm miss}$.

Besides $t \bar t + E_{T, \rm miss}$ also $b \bar b + E_{T, \rm miss}$ production \cite{Lin:2013sca,Artoni:2013zba} has been advocated as a sensitive probe of spin-0 portal couplings to heavy quarks. Recasting the most recent 13~TeV~LHC $b \bar b + E_{T, \rm miss}$ searches \cite{ATLAS-CONF-2016-086,CMS:2016uxr} by means of a simple rescaling  similar to (\ref{eq:monojetratio}) we find that no relevant bound on the parameter space of our simplified model can be derived unless the~$a b \bar b$ coupling is significantly enhanced. From (\ref{eq:xicouplings}) we see that such an enhancement can only arise in THDMs of type II and IV, while it is not possible for the other Yukawa assignments. Since in the limit of large $\tan \beta$ also direct searches for the light pseudoscalar~$a$ in final states containing bottom quarks or charged leptons are relevant (and naively even provide the leading constraints) we do not consider the $b \bar b + E_{T, \rm miss}$ channel in what follows, restricting our numerical analysis to the parameter space with small $\tan \beta$. 
 
\subsection[Mono-$Z$ channel]{Mono-$\bm{Z}$ channel}
\label{sec:monoZ}

\begin{figure}[!t]
\begin{center}
\includegraphics[height=0.225\textwidth]{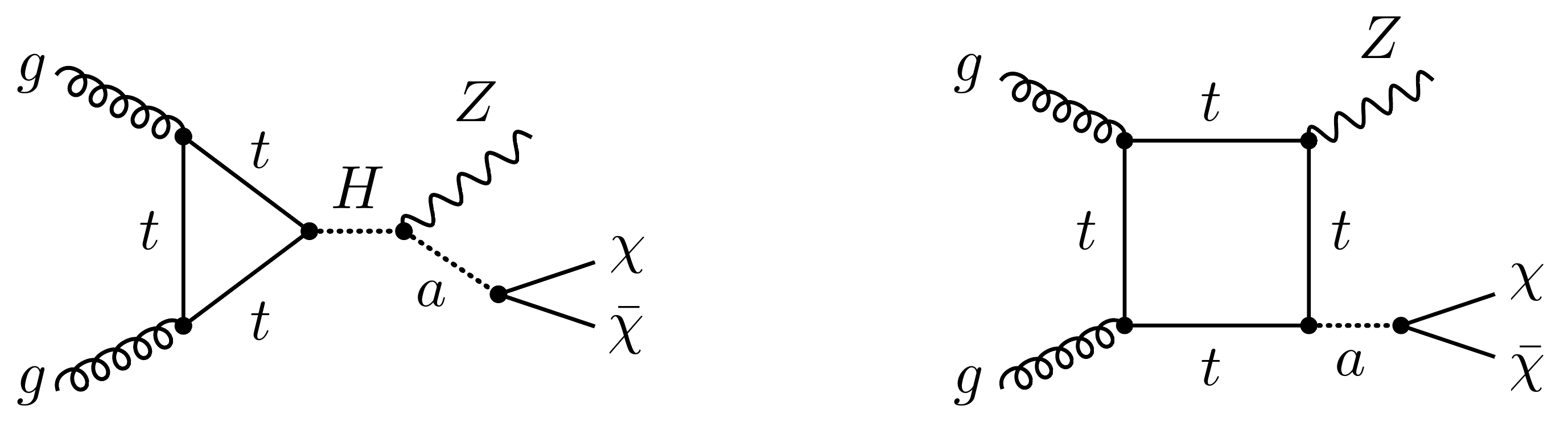}  
\vspace{0mm}
\caption{\label{fig:ZDMDM} Representative Feynman diagrams that lead to a $Z +E_{T,\rm miss}$ signal in the pseudoscalar extensions of the THDM. In the case of triangle diagram (left) only the shown graph contributes, while in the case of the box diagram (right) instead of an $a$  also an~$A$ exchange is possible. }
\end{center}
\end{figure}

A mono-$X$ signal that is strongly suppressed in the case of the spin-0 DMF models~\cite{Mattelaer:2015haa} but will turn out to be relevant in our simplified DM scenario is the mono-$Z$ channel~\cite{Goncalves:2016iyg}. A~sample of one-loop diagrams that lead to such a signature are displayed in Figure~\ref{fig:ZDMDM}. Notice that the left diagram in the figure allows for resonant $Z + \chi \bar \chi$ production through a~$HaZ$ vertex for a sufficiently heavy scalar $H$. Unlike the graph on the right-hand side it has  no counterpart in the spin-0 DMF simplified models. 

As first emphasised in \cite{No:2015xqa} the appearance of the contribution with virtual $H$ and $a$ exchange not only enhances the mono-$Z$ cross section  compared to the spin-0 DMF models, but also leads to quite different kinematics in $Z + \chi \bar \chi$ production. In fact, for  masses $M_H > M_a + M_Z$ the predicted $E_{T , \rm miss}$ spectrum turns out to be peaked at 
\beq \label{eq:ETmissmaxmonoZ}
E_{T, \rm miss}^{\rm max} \simeq \frac{\lambda^{1/2} (M_H, M_a, M_Z)}{2 M_H}  \,,
\eeq
where the two-body phase-space  function $\lambda(m_1, m_2, m_3)$ has been defined in (\ref{eq:lambda}). Denoting the lower experimental requirement on $E_{T, \rm miss}$ in a given mono-$Z$ search by $E_{T, \rm miss}^{\rm cut}$ the latter result can be used to derive a simple bound on $M_H$ for which a significant fraction of the total cross section will pass the cut. We obtain the inequality 
\beq \label{eq:MHETmissineq}
M_H \gtrsim M_a + \sqrt{M_Z^2+ \big( E_{T, \rm miss}^{\rm cut} \big)^2} \,.
\eeq
Given that in the latest mono-$Z$ analyses  \cite{ATLAS:2016bza, CMS:2016hmx, Sirunyan:2017onm}  selection cuts of $E_{T, \rm miss}^{\rm cut} \simeq 100 \, {\rm GeV}$ are imposed it follows that the scalar $H$ has to have a mass of $M_H \simeq 500 \, {\rm GeV}$ if one wants to be sensitive to pseudoscalars $a$ with masses up to the $t \bar t$ threshold $M_a \simeq 350 \, {\rm GeV}$. 

Our detailed Monte Carlo (MC) simulations of the $Z + E_{T, \rm miss}$ signal in Section~\ref{sec:wetdream} however reveals  that the above kinematical argument alone is insufficient to  understand the shape of the mono-$Z$ exclusion in the $M_a$--$\hspace{0.5mm} \tan \beta$ plane in all instances. The reason for this is twofold. First, in cases where $\sin \theta$ is small $H \to aZ$ is often not the dominant~$H$~decay mode and as a result the $Z+ E_{T, \rm miss}$ measurements lose already sensitivity for masses $M_a$ below the bound implied by the estimate (\ref{eq:MHETmissineq}). Second, $Z + \chi \bar \chi$ production in $gg \to a Z$  and  $gg \to A Z$ is also possible through box diagrams, and the interference between triangle and box graphs turns out to  be very relevant  in models that have a light scalar $H$ or pseudoscalar $A$ with a mass below the $t \bar t$ threshold. We add that for $\tan \beta > {\cal O} (10)$ also resonant mono-$Z$ production via $b \bar b \to aZ$ and $b \bar b \to AZ$ can be relevant in models of type~II and IV. In the context of the pure THDM such effects have been studied for instance in~\cite{Harlander:2013mla}.
 
\subsection{Mono-Higgs channel}
\label{sec:monohiggs}

\begin{figure}[!t]
\begin{center}
\includegraphics[height=0.225\textwidth]{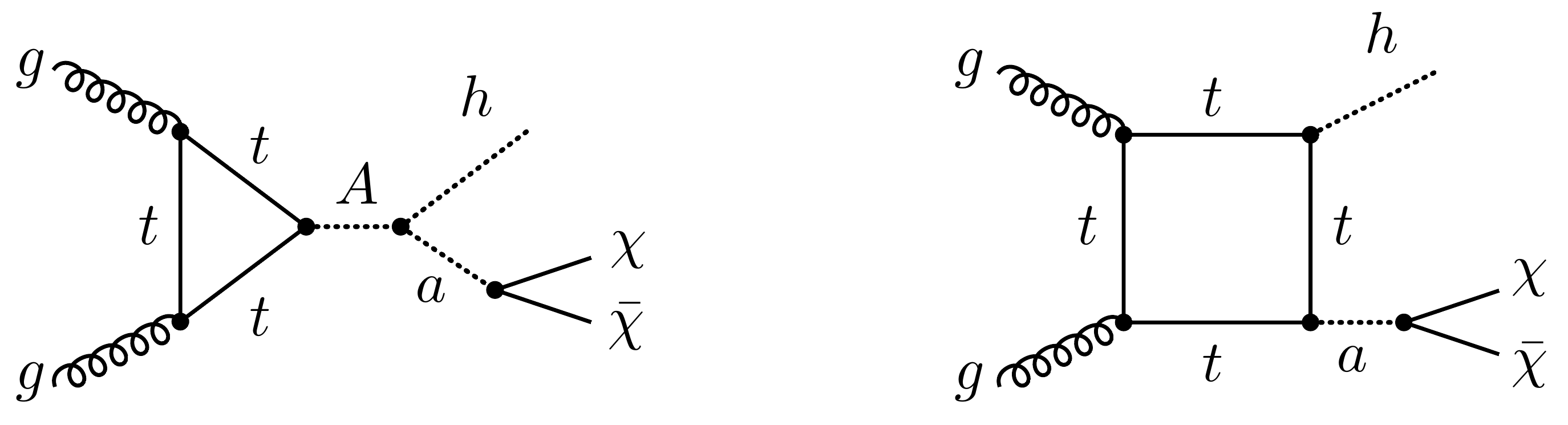}  
\vspace{0mm}
\caption{\label{fig:hDMDM}  Sample diagrams   in the THDM with an extra pseudoscalar that induce a $h +E_{T,\rm miss}$ signal in the alignment/decoupling limit. Graphs in which the role of $a$ and $A$ is interchanged can also provide a relevant contribution.}
\end{center}
\end{figure}

In certain regions of parameter space another possible smoking gun signature of the pseudoscalar extensions of the THDM turns out to be mono-Higgs production. As illustrated in Figure~\ref{fig:hDMDM} this signal can  arise from two different types of one-loop diagrams. For $M_A > M_a + M_h$ the triangle graph with an $Aah$ vertex depicted on the left-hand side allows for resonant mono-Higgs production and  thus dominates over the contribution of the box diagram displayed on the right. In consequence the mono-Higgs production cross sections in the THDM plus pseudoscalar extensions can exceed by far the small  spin-0 DMF model rates for the $h + E_{T, \rm miss}$ signal~\cite{Mattelaer:2015haa}.

Like in the case of the mono-$Z$ signal the presence  of triangle diagrams with a trilinear scalar coupling also leads to a peak in the $E_{T, \rm miss}$ distribution of $h + \chi \bar \chi$ production if the intermediate heavy pseudoscalar $A$ can be resonantly produced. The peak position in the mono-Higgs case is obtained from  \cite{No:2015xqa}
\beq \label{eq:ETmissmaxmonoHiggs}
E_{T, \rm miss}^{\rm max} \simeq \frac{\lambda^{1/2} (M_A, M_a, M_h)}{2 M_A}  \,.
\eeq
It follows that in order for events to pass the $E_{T, \rm miss}$ cut necessary for a background suppression in mono-Higgs searches, the relation 
\beq \label{eq:MAETmissineq}
M_A \gtrsim M_a + \sqrt{M_h^2 + \big( E_{T, \rm miss}^{\rm cut} \big)^2} \,,
\eeq
has to be fulfilled. A lesson to learn from (\ref{eq:MAETmissineq}) is that mono-Higgs searches  in the~$h \to b \bar b$ channel \cite{Aaboud:2016obm, CMS:2016mjh} are less suited to constrain the parameter space of our simplified model than those that focus on $h \to \gamma \gamma$ \cite{ATLAS-CONF-2016-011, CMS:2016xok}, because the minimal~$E_{T, \rm miss}$ requirements in the former analyses are always stricter than those in the latter. To give a relevant numerical example let us consider  $E_{T, \rm miss}^{\rm cut} \simeq 100 \, {\rm GeV}$, which represents a typical~$E_{T, \rm miss}$ cut imposed in the most recent $h + \chi \bar \chi \; (h \to \gamma \gamma)$ searches. From (\ref{eq:MAETmissineq}) one sees that in such a case mono-Higgs analyses are very sensitive to masses up to $M_a \simeq 330 \, {\rm GeV}$ for $M_A \simeq 500 \, {\rm GeV}$. 

Like in the mono-$Z$ case the above  kinematical argument however allows only for a qualitative understanding of the numerical results for the $pp \to h + \chi \bar \chi \; (h \to \gamma \gamma)$ exclusions, since interference effects can be important in scenarios with a pseudoscalar $A$ of mass $M_A <2 m_t$.   Notice that if $M_a > M_A + M_h$ the role of $A$ and $a$ is interchanged and the $h + E_{T, \rm miss}$ signal can receive large corrections from resonant  $a$ exchanges,  as we will see explicitly  in Section~\ref{sec:wetdream}.  Finally  in  type~II and IV  models resonant mono-Higgs production from $b \bar b$ initial states can also be important if $\tan \beta$ is sufficiently large.

\subsection[Mono-$W$ channel]{Mono-$\bm{W}$ channel}
\label{sec:monoW}

\begin{figure}[!t]
\begin{center}
\includegraphics[height=0.165\textwidth]{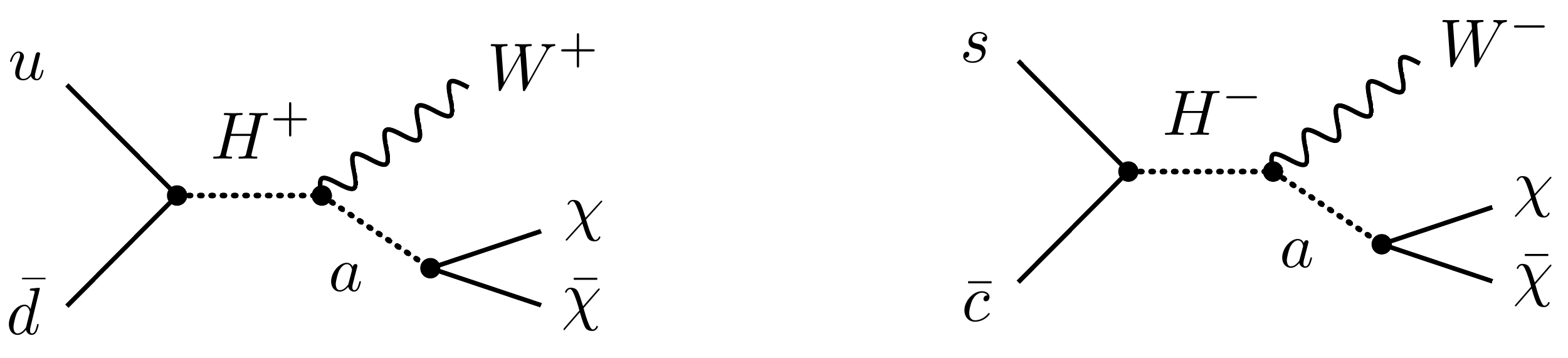}  
\vspace{0mm}
\caption{\label{fig:WDMDM} Examples of  diagrams that lead to  a $W +E_{T,\rm miss}$ signature through the exchange of a charged Higgs $H^\pm$ and a lighter pseudoscalar $a$ in the THDM plus pseudoscalar extension.  }
\end{center}
\end{figure}

The last $E_{T, \rm miss}$ signal that we consider is the mono-$W$ channel~\cite{Aaboud:2016zkn,Khachatryan:2016jww}. Two representative Feynman graphs that lead to a resonant $W + E_{T, \rm miss}$ signature in the pseudoscalar extension of the THDM are shown in Figure~\ref{fig:WDMDM}. These diagrams describe the single production of a charged Higgs $H^\pm$ via the annihilation of light quarks followed by $H^\pm \to a W^\pm \, (a \to \chi \bar \chi)$. One way to assess the prospects for detecting a mono-$W$ signature consists in comparing the production cross sections of $H^\pm$ to that of $H$ and $A$. Using for instance $\tan \beta = 1$, we find $\sigma \left ( pp \to H^+ \right ) \simeq 1.0 \, {\rm fb}$ for $M_{H^\pm} = 500 \, {\rm GeV}$ and $\sigma \left ( pp \to H^+ \right ) \simeq 0.2 \, {\rm fb}$ for $M_{H^\pm} = 750 \, {\rm GeV}$ at the 13 TeV LHC. The corresponding cross sections in the case of the heavy  neutral spin-0 resonances read $\sigma \left ( pp \to H \right ) \simeq 1.4 \, {\rm pb}$ and $\sigma \left ( pp \to A \right ) \simeq 3.1 \, {\rm pb}$ and $\sigma \left ( pp \to H \right ) \simeq 0.2 \, {\rm pb}$ and $\sigma \left ( pp \to A \right ) \simeq 0.3 \, {\rm pb}$, respectively. These numbers strongly suggest that an observation of a mono-$W$ signal is compared to that of a mono-$Z$ or mono-Higgs signature much less probable. We thus do not consider the $W + E_{T, \rm miss} $ channel any further. 

Let us finally add that besides a simple mono-$W$ signature also $Wt + E_{T, \rm miss}$ and $Wtb+E_{T, \rm miss}$ signals can appear in the DM model introduced in Section~\ref{sec:THDMP}.  For the relevant charged Higgs  production cross sections we find at 13~TeV the results $\sigma \left ( g \bar b \to H^+ \bar t \right ) \simeq 0.17 \, {\rm pb}$ ($\sigma \left ( g \bar b \to H^+ \bar t \right ) \simeq 0.04 \, {\rm pb}$) and $\sigma \left ( g g \to H^+  \bar t  b\right ) \simeq 0.10 \, {\rm pb}$ ($\sigma \left ( gg \to H^+ \bar t b \right ) \simeq 0.02 \, {\rm pb}$) using $\tan \beta =1$ and $M_{H^\pm} = 500 \, {\rm GeV}$ ($M_{H^\pm} = 750 \, {\rm GeV}$). Given the small  $H^\pm$  production cross section in $gb$ and $gg$ fusion, we expect that searches for a $Wt + E_{T, \rm miss}$ or a $Wtb + E_{T, \rm miss}$ signal will  in practice provide no relevant constraint in the small $\tan \beta$ regime. 

\section{Numerical results}
\label{sec:numerics}

The numerical results of our mono-$X$ analyses are presented in this section. After a brief description of the signal generation and the background estimates, we first study the impact of interference effects between the $a$ and $A$ contributions to  the $j + \chi \bar \chi$ and $t \bar t + \chi \bar \chi$ channels. Then the constraints on the parameter space of the THDM plus pseudoscalar extensions  are derived for several well-motivated benchmark scenarios. In the case of the mono-$Z$ and mono-Higgs searches we also discuss the LHC Run II reach in some detail.  

\subsection{Signal generation}
\label{sec:signalgeneration}

The starting point of our MC simulations is a  {\tt UFO} implementation~\cite{Degrande:2011ua} of the simplified model as described in Section~\ref{sec:THDMP}. This implementation has been obtained by means of the {\tt FeynRules~2}~\cite{Alloul:2013bka} and  {\tt NLOCT}~\cite{Degrande:2014vpa} packages. The generation of the $j + E_{T, \rm miss}$, $Z + E_{T, \rm miss} \, (Z \to \ell^+ \ell^-)$ and $h + E_{T, \rm miss} \, (h \to \gamma \gamma)$ signal samples is performed at leading order~(LO) with {\tt MadGraph5\_aMC@NLO}~\cite{Alwall:2014hca} using~{\tt PYTHIA~8.2} \cite{Sjostrand:2014zea} for showering and {\tt NNPDF2.3}~\cite{Ball:2012cx} as parton distribution functions. The whole MC chain is steered with {\tt CheckMATE~2}~\cite{Dercks:2016npn} which itself employs {\tt FastJet}~\cite{Cacciari:2011ma} to reconstruct hadronic jets and {\tt Delphes~3}~\cite{deFavereau:2013fsa} as a fast-detector simulation. The results of the  {\tt CheckMATE~2} analyses have been validated against {\tt MadAnalysis~5}~\cite{Conte:2012fm,Conte:2014zja}.  The selection requirements imposed in our analyses resemble those used in the recent  LHC mono-jet~\cite{Aaboud:2016tnv}, mono-$Z$~\cite{ATLAS:2016bza}  and mono-Higgs~\cite{CMS:2016xok}  search, respectively. For what concerns our $t \bar t + E_{T, \rm miss} \, (t \to  \ell b \nu)$ recast we rely on the results of the sensitivity study \cite{Haisch:2016gry}.  In this analysis the DM signal has been simulated at next-to-leading order (NLO) with {\tt MadGraph5\_aMC@NLO} and {\tt PYTHIA~8.2} using a {\tt FxFx} NLO jet matching prescription~\cite{Frederix:2012ps} and the final-state top quarks have been decayed with {\tt MadSpin}~\cite{Artoisenet:2012st}.

\subsection{Background estimates}
\label{sec:backgroundestimates}

For the $j + E_{T, \rm miss}$, $Z + E_{T, \rm miss} \, (Z \to \ell^+ \ell^-)$ and $h + E_{T, \rm miss} \, (h \to \gamma \gamma)$  recasts our background estimates rely on the background predictions obtained in the 13~TeV~LHC analyses~\cite{Aaboud:2016tnv},~\cite{ATLAS:2016bza}  and~\cite{CMS:2016xok}, respectively.  The given background numbers correspond to  $3.2 \, {\rm fb}^{-1}$, $13.3 \, {\rm fb}^{-1}$, $2.3 \, {\rm fb}^{-1}$  and we extrapolate them to $40 \, {\rm fb}^{-1}$ of integrated luminosity to be able to assess the near-term reach of the different mono-$X$ channels. Our extrapolations assume that while the relative systematic uncertainties remain the same, the relative statistical errors scale  as~$1/\sqrt{L}$ with luminosity $L$. Depending on the signal region the relative systematic uncertainties amount to around $4\%$ to $9\%$ in the case of the mono-jet search, about~$7\%$ for the mono-$Z$ analysis and approximately~$20\%$ for the mono-Higgs channel. 

Since the $ j + E_{T, \rm miss}$ search is already systematics limited at~$40 \, {\rm fb}^{-1}$ its constraining power  will depend sensitively on the assumption about the future systematic uncertainty on the associated SM background. This  should be kept  in mind when comparing the different exclusions presented below, because a better understanding of the backgrounds can have a visible impact on the obtained results. Since the $t \bar t + E_{T, \rm miss} \, (t \to  \ell b \nu)$ search will still be statistically limited for  $40 \, {\rm fb}^{-1}$, we  base our forecast in this case  on a data set of  $300 \, {\rm fb}^{-1}$  assuming that the relevant SM background is known to 20\%. In the mono-$Z$ and mono-Higgs cases we will  present below, besides $40 \, {\rm fb}^{-1}$ projections, results for $100 \, {\rm fb}^{-1}$ and $300 \, {\rm fb}^{-1}$ of data. From these results one can assess if the existing $Z + E_{T, \rm miss}$ and $h + E_{T, \rm miss}$ search strategies will at some point become systematics limited  in LHC Run II.

\subsection{Interference effects}
\label{sec:interferences}

Our simplified model contains two pseudoscalar mediators $a$ and $A$ that are admixtures of the  neutral CP-odd weak eigenstates entering  (\ref{eq:VH}) and (\ref{eq:VP}). In mono-jet production the two contributions interfere and the resulting  LO matrix element  takes the following schematic form
\beq \label{eq:interference}
{\cal M} \left ( p p \to j + \chi \bar \chi \right ) \propto \frac{1}{m_{\chi \bar \chi}^2 -M_a^2 - i M_a \Gamma_a } - \frac{1}{m_{\chi \bar \chi}^2 -M_A^2 - i M_A \Gamma_A} \,,
\eeq
where $m_{\chi \bar \chi}$ denotes the invariant mass of the DM pair and $\Gamma_a$ and $\Gamma_A$ are  the total decay widths of the two pseudoscalar mass eigenstates.
The same results hold for instance also in the case of the $pp \to t \bar t +  \chi \bar \chi$ amplitude. Notice that the contributions from virtual~$a$ and $A$ exchange have opposite signs in (\ref{eq:interference}) resulting from the transformation from the weak to the mass eigenstate basis. Such a destructive interference of  two contributions also appears in fermion scalar singlet models with Higgs mixing and has there  shown to be phenomenologically relevant~\cite{Kim:2008pp,Baek:2011aa,LopezHonorez:2012kv,Ko:2016ybp,Baek:2017vzd}. 

\begin{figure}[!t]
\begin{center}
\includegraphics[width=0.425\textwidth]{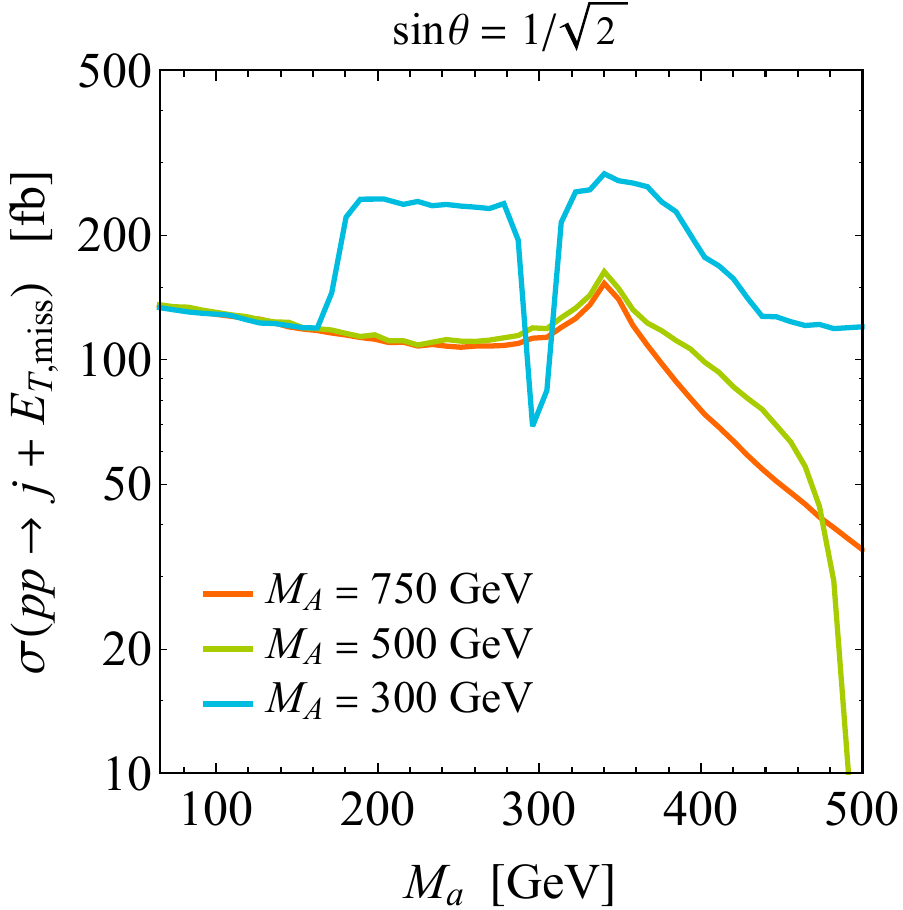}  \qquad 
\includegraphics[width=0.425\textwidth]{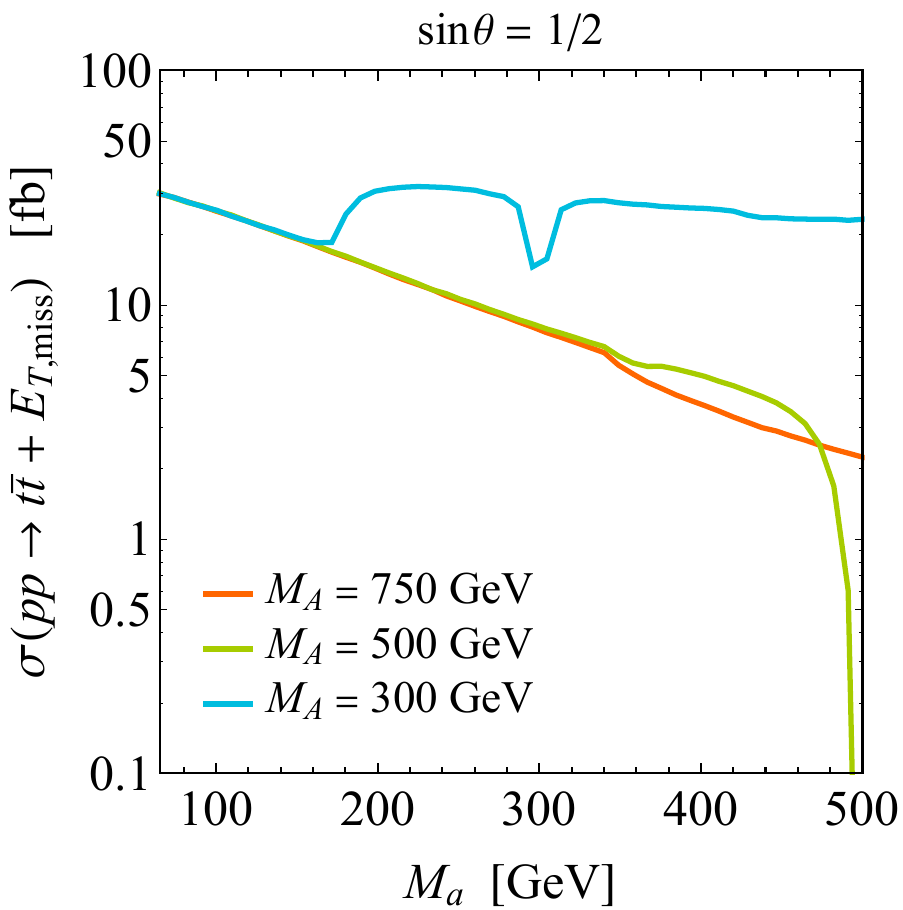} 
\vspace{0mm}
\caption{\label{fig:interferences} Predictions for the mono-jet ($t \bar t + E_{T, \rm miss}$) cross section as a function of $M_a$ for three different values of $M_A$. In the left (right) plot $\sin \theta = 1/\sqrt{2}$ ($\sin \theta = 1/2$) is used and the other relevant parameters are  $\tan \beta = 1$, $M_H = M_{H^\pm} = 750 \, {\rm GeV}$, $\lambda_3 = \lambda_{P1} = \lambda_{P2} = 0$, $y_\chi = 1$ and $m_\chi = 1 \, {\rm GeV}$.  The shown results correspond to~13~TeV $pp$ collisions and employ minimal sets of cuts as explained in the main text. }
\end{center}
\end{figure}

The impact of interference effects on the predictions of the mono-jet and $t \bar t + E_{T, \rm miss}$ cross sections is illustrated in Figure~\ref{fig:interferences} for three different values of the mass of the pseudoscalar~$A$.  Both plots display partonic LO results at 13~TeV LHC energies.  In the left  panel the basic selection requirements $E_{T, \rm miss} > 250 \, {\rm GeV}$ and $|\eta_j| < 2.4$  are imposed with $\eta_j$ denoting the pseudorapidity of the jet, while in the right figure only the cut $E_{T, \rm miss} > 150 \, {\rm GeV}$ is applied. Focusing first on  the cross sections for $M_A = 750 \, {\rm GeV}$~(red curves), one observes that in this case interference effects do not play any role since the pseudoscalar $A$ is too heavy and effectively decouples. One also sees  that at $M_a \simeq 350 \, {\rm GeV}$ the cross sections of both mono-jet and $t \bar t + E_{T, \rm miss}$ production are enhanced due to $t \bar t$ threshold effects. Notice furthermore that the enhancement is more pronounced for the $j + E_{T, \rm miss}$ signal because the top-quark loops develop an imaginary piece once the internal tops can go on-shell. 

The results for $M_A = 500 \, {\rm GeV}$ (green curves) resemble closely those for $M_A = 750 \, {\rm GeV}$ until $M_a \simeq M_A - M_h \simeq 375 \, {\rm GeV}$ at which point one observes an enhancement of the rates compared to the case of very heavy $A$. This feature is a consequence of  the fact that for $M_a < M_A - M_h$ the $A \to ah$ channel is the dominant decay mode of~$A$, as can be seen from the right plot in Figure~\ref{fig:BrA}. For larger masses of $a$ the phase space of $A \to ah$ closes  and in turn ${\rm BR} \left (A \to \chi \bar \chi \right )$ increases. This leads to  constructive interference between the two terms in (\ref{eq:interference}) until $M_a \simeq M_A = 500 \, {\rm GeV}$ where the interference becomes destructive. Notice furthermore that the same qualitative explanations apply to the case of $M_A = 300 \, {\rm GeV}$ (blue curves) where the constructive and destructive interference takes place at $M_a \simeq M_A - M_h \simeq 175 \, {\rm GeV}$ and $M_a \simeq M_A = 300 \, {\rm GeV}$, respectively. Comparing the left and right panel of Figure~\ref{fig:interferences}, one finally sees that the observed interference pattern is at the qualitative level independent of the choice of $\sin \theta$. 

\subsection{Summary plots}
\label{sec:wetdream}

Below we study four different benchmark scenarios that exemplify the rich~$E_{T,\rm miss}$ phenomenology of the simplified DM model introduced in Section~\ref{sec:THDMP}. Throughout our analysis we work in the alignment/decoupling limit, adopting the parameters $M_{H^\pm} = 750 \, {\rm GeV}$, $\lambda_3 = \lambda_{P1} = \lambda_{P2} = 0$, $y_\chi = 1$ and $m_\chi= 1 \, {\rm GeV}$ and consider a Yukawa sector of type~II.  The shown results however also hold in the case of the other Yukawa sectors (\ref{eq:yukawatypes}) since for $\tan \beta = {\cal O} (1)$ effects of bottom-quark loops in mono-jet, mono-$Z$ and mono-Higgs production amount to corrections  of a few percent only. The model-dependent contributions from $b \bar b$-initiated production also turn  out to be small for such values of $\tan \beta$. The constraints on all benchmark scenarios will be presented in the $M_a$--$\hspace{0.5mm} \tan \beta$ plane, in which  the  parameter regions that are excluded at 95\%~CL  by the various searches will be indicated.

\begin{figure}[!t]
\begin{center}
\includegraphics[width=0.425\textwidth]{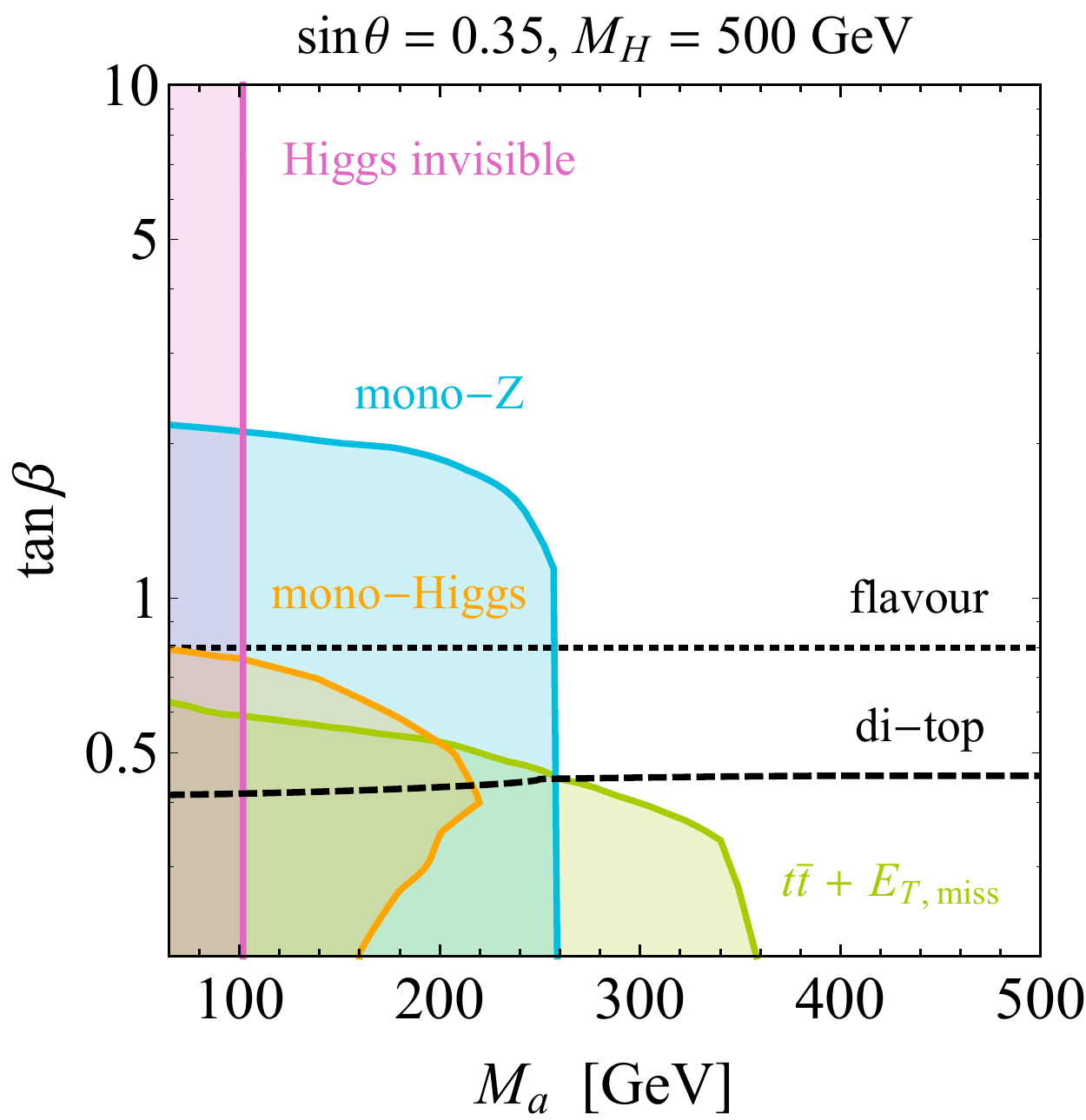}  \qquad 
\includegraphics[width=0.425\textwidth]{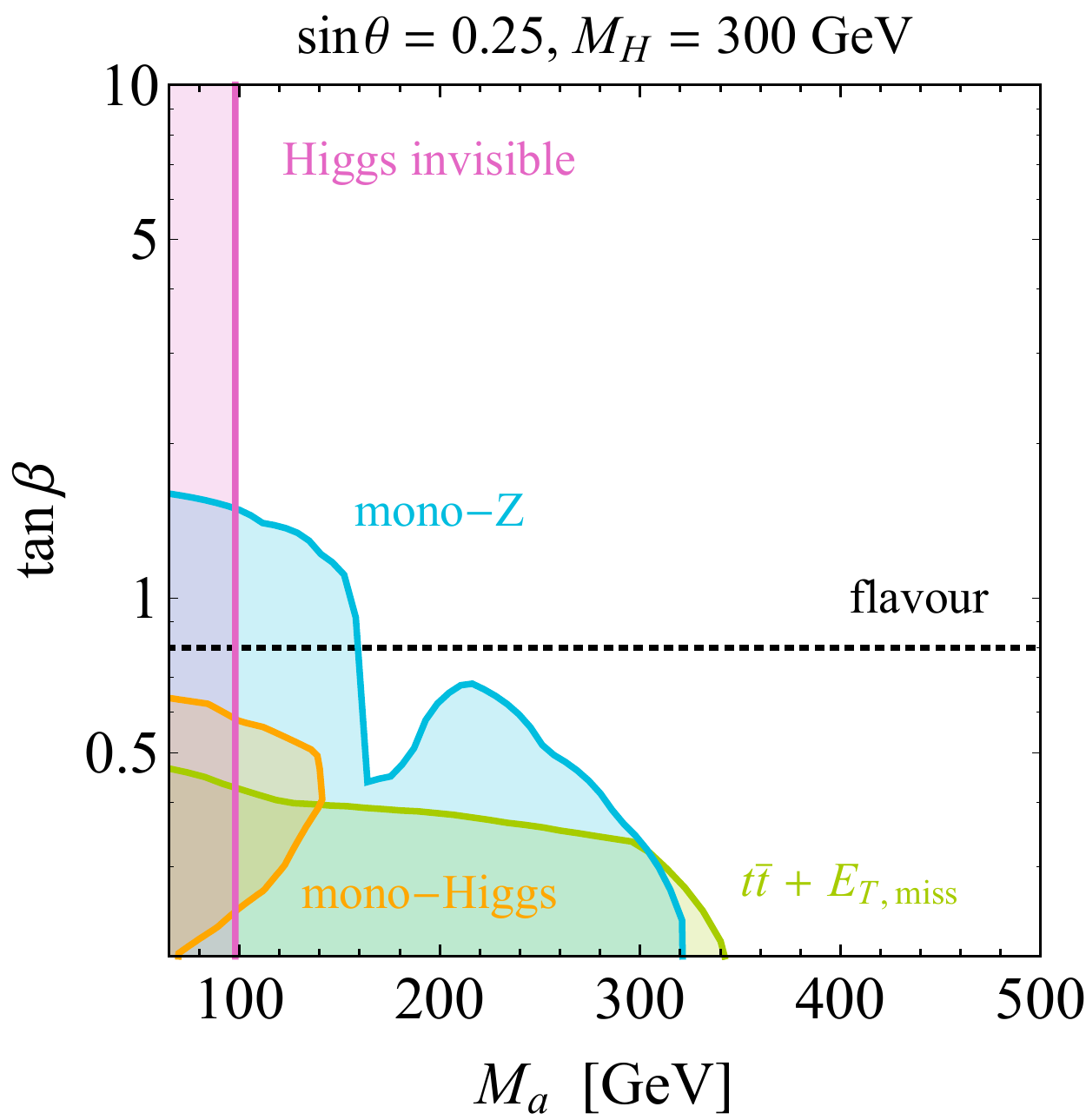} 

\vspace{5mm}

\includegraphics[width=0.425\textwidth]{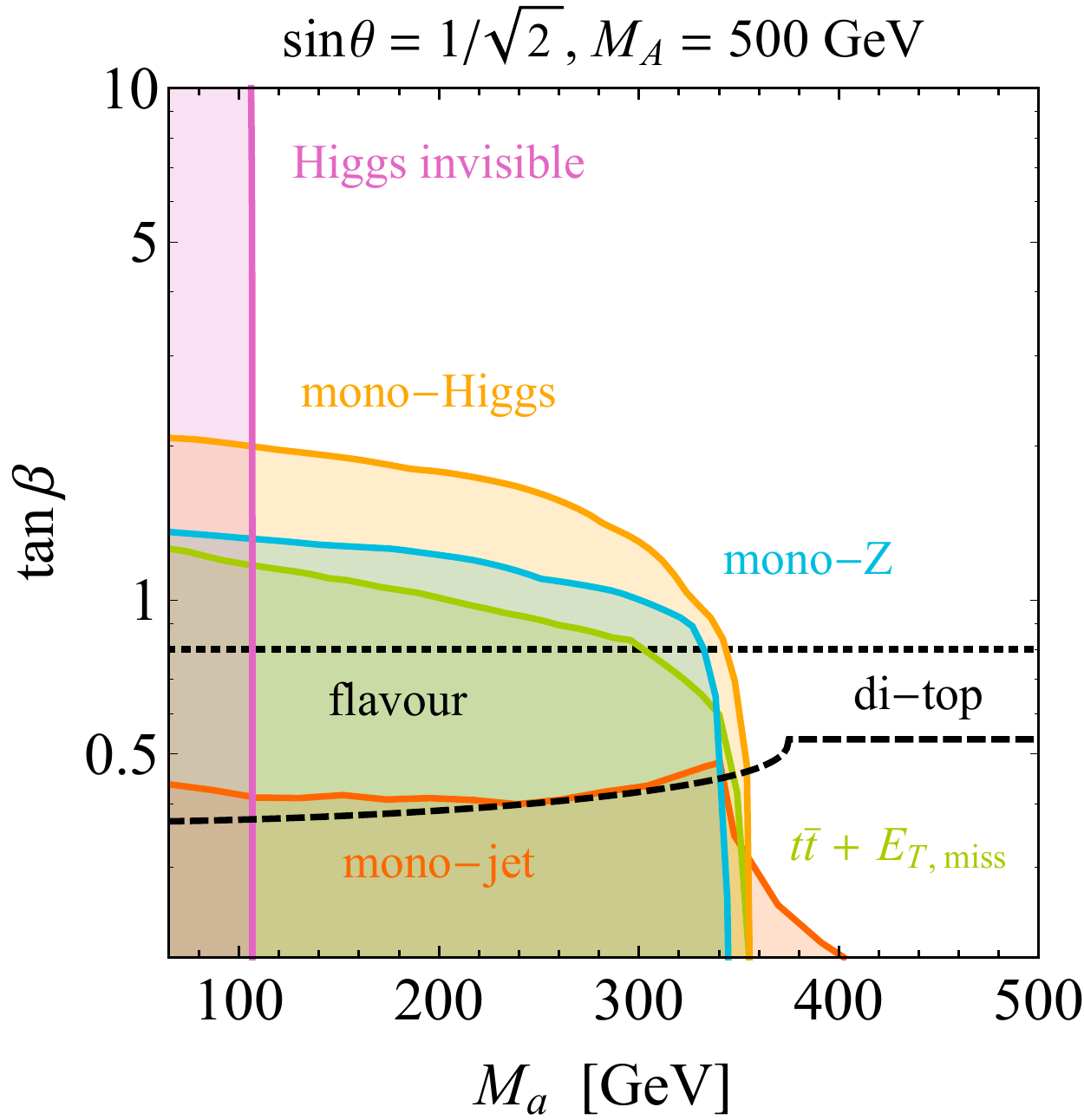}  \qquad 
\includegraphics[width=0.425\textwidth]{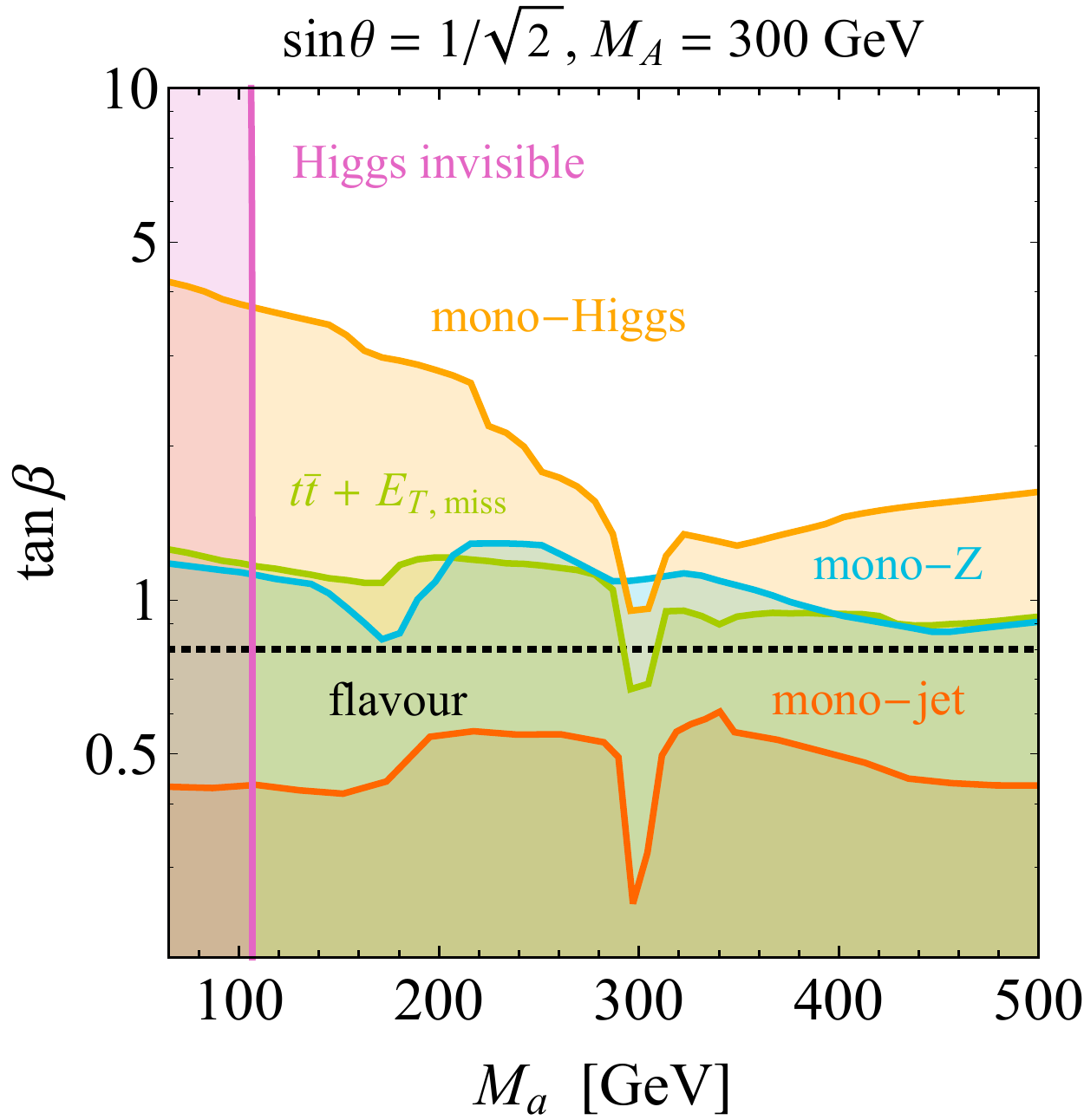} 
\vspace{0mm}
\caption{\label{fig:wetdream} Summary plots showing all relevant constraints in the $M_a$--$\hspace{0.5mm} \tan \beta$ plane for four  benchmark scenarios. The colour shaded regions correspond to the parameter space excluded by the different $E_{T, \rm miss}$ searches, while the constraints arising from di-top resonance searches and flavour physics are indicated by the dashed and dotted black lines, respectively. Parameters choices below the black lines are excluded. All exclusions are 95\%~CL bounds. See text for further details.}
\end{center}
\end{figure}

\subsubsection*{Benchmark scenario 1: $\bm{\sin \theta = 0.35}$, $\bm{M_H = 500 \, {\rm GeV}}$}

In the first benchmark scenario we choose $\sin \theta = 0.35$, $M_H = 500 \, {\rm GeV}$ and  $M_A = 750 \, {\rm GeV}$, where the choice of $\sin \theta$ guarantees that EW precision measurements are satisfied for  all  values of $M_a$ that we consider (see Section~\ref{sec:EWprecision}). The upper left panel in Figure~\ref{fig:wetdream} summarises the various 95\%~CL exclusions. One first observes that the constraint from invisible decays of the Higgs (pink region) excludes all shown values of $\tan \beta$ for mediator masses of $M_a \lesssim 100 \, {\rm GeV}$. This constraint has been obtained by imposing the 95\%~CL limit ${\rm BR} \left ( h \to {\rm invisible} \right ) < 25\%$ set  by ATLAS~\cite{Aad:2015pla}. Notice that in the THDM plus pseudoscalar extensions one has ${\rm BR} \left ( h \to {\rm invisible} \right ) \simeq {\rm BR} \left ( h \to 2 \chi 2 \bar \chi \right ) \simeq 100\%$ for a~DM mass of $m_\chi = 1 \, {\rm GeV}$ largely independent of $\sin \theta$, $M_H$ and $M_A$, and as a result the $h \to {\rm invisible}$ constraint is roughly the same in all of our benchmark scenarios. One furthermore sees that taken together the existing limits  from flavour physics~(dotted black line) and di-top searches~(dashed black curve)  exclude the parameter region with $\tan \beta \lesssim 0.8$. Here the di-top constraint is obtained from the results~\cite{ATLAS:2016pyq} by rescaling the limit quoted by ATLAS using the $t \bar t$ branching ratio of the heavy scalar mediator $H$ (see Section~\ref{sec:ditop}). 

Turning ones attention to the constraints that arise from DM searches, one observes that even with an integrated luminosity of $300 \, {\rm fb}^{-1}$, $t \bar t + E_{T, \rm miss}$ measurements~(green region) should be able to exclude only a small part of the $M_a$--$\hspace{0.5mm} \tan \beta$ plane. For pseudoscalar masses~$M_a$ around the EW scale values of $\tan \beta \lesssim 0.6$  can be tested, while $t \bar t + E_{T, \rm miss}$ searches have essentially no sensitivity to the parameter region with $M_a \gtrsim 2 m_t$  since the decay channel $a \to t \bar t$ opens up. The weakness of the $t \bar t + E_{T, \rm miss}$ constraint  is expected~$\big ($see~(\ref{eq:monojetratio})$\big)$ since  the $t \bar t + a$ production cross section is suppressed by $\sin^2 \theta \simeq 0.1$ in our first benchmark. This suppression is also the reason for our finding that with $40 \, {\rm fb}^{-1}$ of 13~TeV data, mono-jet searches will not lead to any relevant restriction on $\tan \beta$, if one assumes that these near-future measurements are plagued by systematic uncertainties at the 5\% level in the low-$E_{T, \rm miss}$ signal regions.  

The hypothetical mono-$Z$ search (blue region) based on~$40 \, {\rm fb}^{-1}$ of data provides the strongest  constraint for $M_a \lesssim 250 \, {\rm GeV}$, excluding $\tan \beta$ values slightly above~2 for light mediators $a$. This strong bound is a result of the resonant enhancement of $Z + \chi \bar \chi$ production in our first benchmark scenario. Notice furthermore the sharp cut-off  of the $Z + E_{T, \rm miss}$ exclusion at $M_a \simeq 260 \, {\rm GeV}$. For larger pseudoscalar masses $M_a$ one finds that ${\rm BR} \left (H \to aZ \right ) \lesssim 10\%$  (see the right panel in Figure~\ref{fig:BrH}) and as a result  mono-$Z$ production through triangle graphs is strongly reduced. This explains why the $Z + E_{T, \rm miss}$ search looses sensitivity already before $M_a \simeq 350 \, {\rm GeV}$ as one would naively expect from~(\ref{eq:MHETmissineq}) for the $E_{T, \rm miss}^{\rm cut} = 120 \, {\rm GeV}$  high-mass signal region  requirement imposed in \cite{ATLAS:2016bza}. We finally see that with $40 \, {\rm fb}^{-1}$ of integrated luminosity mono-Higgs searches~(orange region) can  cover only a small part of the parameter space compared to mono-$Z$ measurements. 

\subsubsection*{Benchmark scenario 2: $\bm{\sin \theta = 0.25}$, $\bm{M_H = 300 \, {\rm GeV}}$}

In our second benchmark scenario, the sine of the mixing angle is $\sin \theta =0.25$ and the masses of $H$ and $A$ are taken to be $M_H = 300 \, {\rm GeV}$ and  $M_A = 750 \, {\rm GeV}$. The corresponding exclusion contours are depicted in the upper right panel of  Figure~\ref{fig:wetdream}. The constraints from $h \to {\rm invisible}$~(pink region) and flavour physics (dotted black line) resemble the exclusions that apply in the first benchmark case. The recent ATLAS di-top search does instead not lead to a constraint since, on the one hand, $t \bar t$~decays of the scalar $H$ are kinematically forbidden, and on the other hand, the  ATLAS sensitivity to very heavy pseudoscalars~$A$ is not sufficient to set a bound on $\tan \beta$.  

Given the smallness of $\sin \theta$, we find that our hypothetical   $t \bar t + E_{T, \rm miss}$ search  only probes the parameter region with $M_a \lesssim 2m_t$ and $\tan \beta \lesssim 0.4$. Mono-jet measurements are expected to provide even weaker restrictions and in consequence  we do not show  the corresponding bounds. As in the case of the first benchmark scenario, the mono-$Z$ exclusion~(blue region) is the most stringent constraint for a large range of $M_a$ values, excluding values of $\tan \beta \lesssim 1.5$ for  $M_a \simeq M_h$. The dip of the exclusion limit at $M_a \simeq 170 \, {\rm GeV}$ coincides with the bound derived in (\ref{eq:MHETmissineq}) if the low-mass signal region requirement $E_{T, \rm miss}^{\rm cut} = 90 \, {\rm GeV}$~\cite{ATLAS:2016bza} is imposed.  One also observes that for larger mediator masses the mono-$Z$ exclusion strengthens until the point where $M_a \simeq 220 \, {\rm GeV}$. This is a result of the constructive interference  between triangle and box graphs (see Figure~\ref{fig:ZDMDM}). The bound that follows from our  $40 \, {\rm fb}^{-1}$ mono-Higgs projection (orange region) is compared to the mono-$Z$ exclusion again rather weak. 

\subsubsection*{Benchmark scenario 3: $\bm{\sin \theta = 1/\sqrt{2}}$, $\bm{M_A = 500 \, {\rm GeV}}$}

Our third benchmark scenario employs $\sin \theta = 1/\sqrt{2}$, $M_A = 500 \, {\rm GeV}$ and  $M_H  = 750 \, {\rm GeV}$. Notice that for $M_H = M_{H^\pm}$  the mixing in the pseudoscalar sector can be  large since there are no constraints on $\sin \theta$ and $M_a$ from $\Delta \rho$. The  constraints on the $M_a$--$\hspace{0.5mm} \tan \beta$ plane corresponding to these parameter choices  are presented in the lower left panel of~Figure~\ref{fig:wetdream}. The bounds from $h \to {\rm invisible}$ decays~(pink region) and flavour physics~(dotted black line) are essentially unchanged with respect to the previous benchmarks. The shown di-top constraint (dashed black curve) differs from the one displayed in the upper left panel since it follows from the bound  provided in~\cite{ATLAS:2016pyq} for a pseudoscalar with a mass of $500 \, {\rm GeV}$. 

In the case of the mono-jet constraint~(red region) one sees that it should now be possible to exclude $\tan \beta \lesssim 0.4$ values for $M_a \lesssim 350 \, {\rm GeV}$. One furthermore observes that future $t \bar t + E_{T, \rm miss}$ searches~(green region) are expected to extend the parameter space excluded by the non-$E_{T, \rm miss}$ constraints to~$\tan \beta$ values above 1  for  $M_a \lesssim 200 \, {\rm GeV}$. Although the scalar~$H$ is very heavy, we find that the  mono-$Z$ projection~(blue region) still provides relevant constraints in the  $M_a$--$\hspace{0.5mm} \tan \beta$ for masses below the~$a \to t \bar t$ threshold, because the mixing angle $\theta$ is maximal in our third benchmark. The strongest~$E_{T, \rm miss}$ constraint  is however provided by the mono-Higgs search~(orange region),  which should be able to exclude values of~$\tan \beta \lesssim 2$ for pseudoscalars~$a$ with masses at the EW scale. Notice that the mono-Higgs exclusion has a sharp cut-off at $M_a \simeq 350 \, {\rm GeV}$, as expect from the inequality (\ref{eq:MAETmissineq}) for $E_{T, \rm miss}^{\rm cut} = 105 \, {\rm GeV}$~\cite{CMS:2016xok}. 

\subsubsection*{Benchmark scenario 4: $\bm{\sin \theta = 1/\sqrt{2}}$, $\bm{M_A = 300 \, {\rm GeV}}$}

In the fourth benchmark we consider the parameters $\sin \theta = 1/\sqrt{2}$, $M_A = 300 \, {\rm GeV}$ and  $M_H =  750 \, {\rm GeV}$. As can be seen from the lower right panel of Figure~\ref{fig:wetdream}, the regions excluded by  Higgs to invisible decays~(pink region) and flavour physics (dotted black line) are close to identical to those arising in all the other  scenarios. In contrast, di-top searches do not lead to a restriction because the pseudoscalar $A$ is too light to decay to two on-shell top quarks, while the ATLAS search~\cite{ATLAS:2016pyq} is not yet sensitive to very heavy scalars $H$. 

The shapes of the exclusions from the $j + E_{T, \rm miss}$~(red region) and $t \bar t + E_{T, \rm miss}$~(green region) measurements display an interference pattern that is very similar to the one seen in~Figure~\ref{fig:interferences}. In turn  future mono-jet ($t \bar t + E_{T, \rm miss}$) searches are expected to be able to exclude $\tan \beta \lesssim 0.4$  ($\tan \beta \lesssim 1$) values for mediator masses $M_a$ above the $t \bar t$ threshold. Focusing  our attention on the mono-$Z$ projection~(blue region) we observe that the corresponding exclusion curve has a pronounced dip at $M_a \simeq 180 \, {\rm GeV}$. It originates from the interference of triangle diagrams with box graphs  that correspond to $gg \to AZ \to Z+\chi \bar \chi$ (see Figure~\ref{fig:ZDMDM}). This interference is destructive and maximal when the decay channel $A \to aZ$ starts to close,  leading to ${\rm Br} \left (A \to \chi \bar \chi \right ) \simeq 100\%$ for the considered value of~$M_A$. 

 Like in the third benchmark  the mono-Higgs search~(orange region) is again the most powerful $E_{T, \rm miss}$ constraint as it allows to exclude $\tan \beta \lesssim 3.7$ values for $M_a \simeq 100 \, {\rm GeV}$. We also note that the mono-Higgs search maintains sensitivity for $M_a$ values well above the estimate presented in~(\ref{eq:MAETmissineq}). The reason is that for sufficiently light pseudoscalars $A$,  triangle diagrams with resonant $a$ exchange (see Figure~\ref{fig:hDMDM}) can provide a sizeable contribution to mono-Higgs production. This resonant enhancement allows one to probe values of $\tan \beta$ above 1 for $M_a \gtrsim 300 \, {\rm GeV}$. Notice finally that at $M_a \simeq M_A  = 300 \, {\rm GeV}$ the $a$ and $A$  contributions  interfere destructively leading to a visible dip in the $h + E_{T, \rm miss}$ exclusion.  

\subsection{LHC Run II reach}
\label{sec:prospects}

\begin{figure}[!t]
\begin{center}
\includegraphics[width=0.425\textwidth]{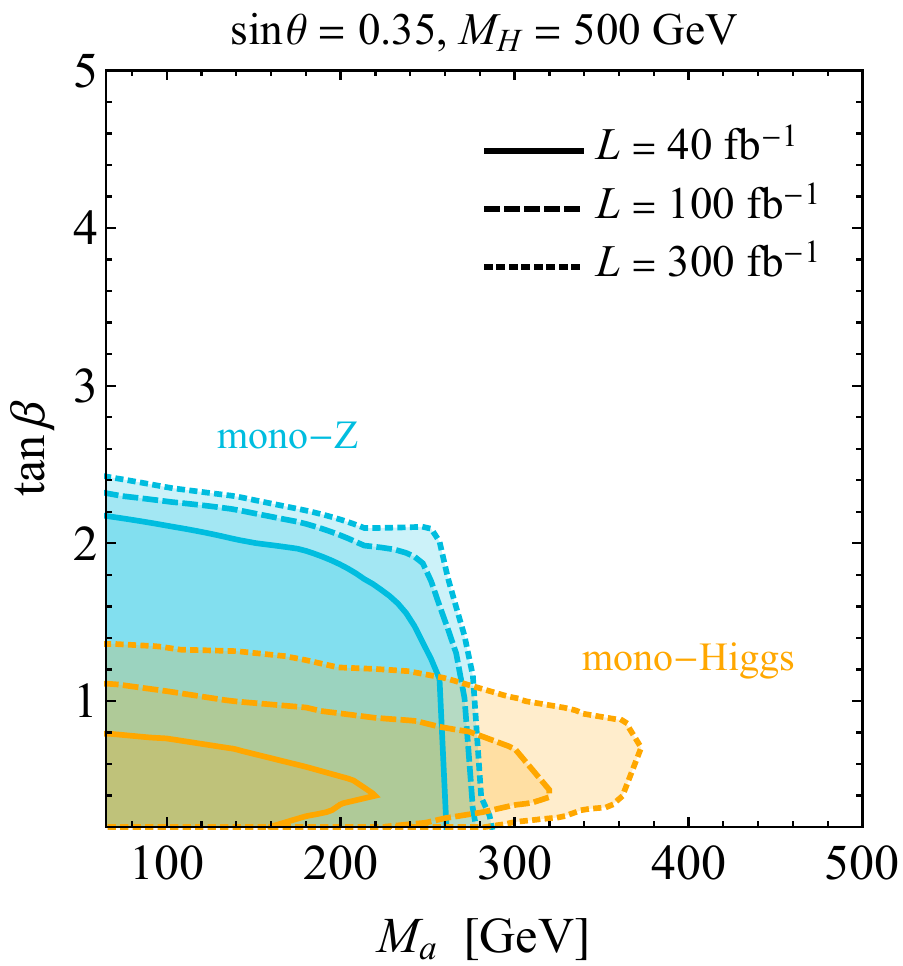}  \qquad 
\includegraphics[width=0.425\textwidth]{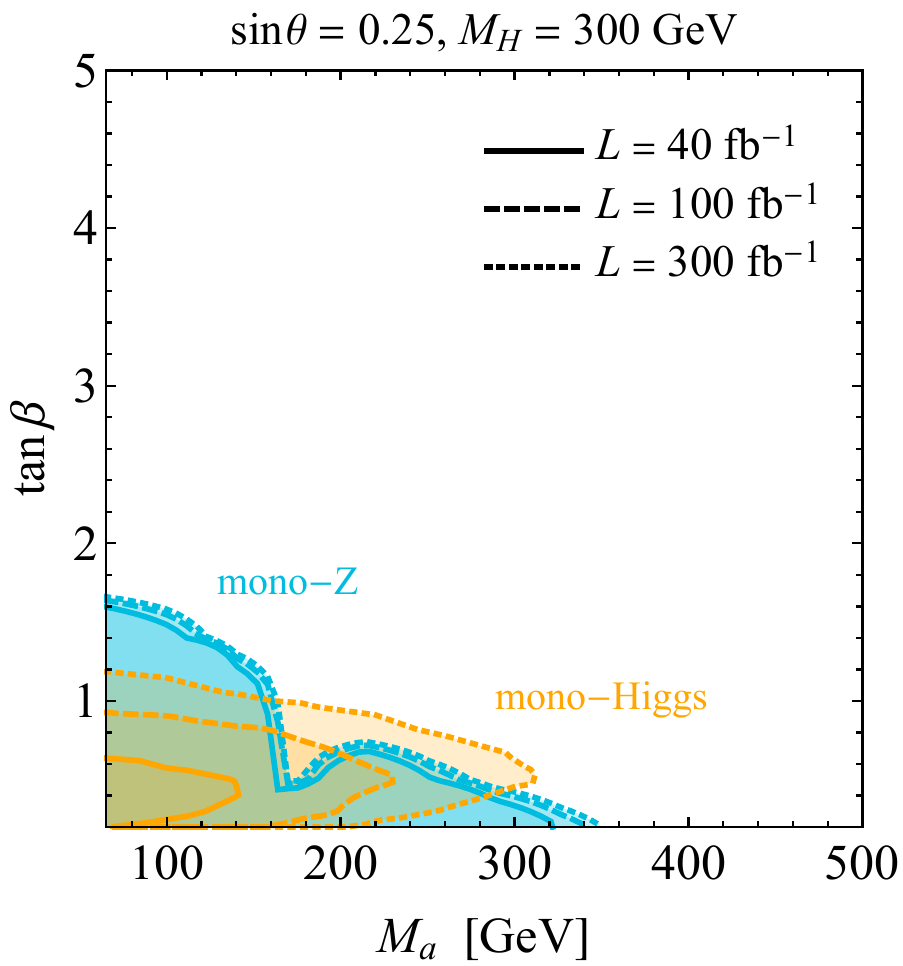} 

\vspace{5mm}

\includegraphics[width=0.425\textwidth]{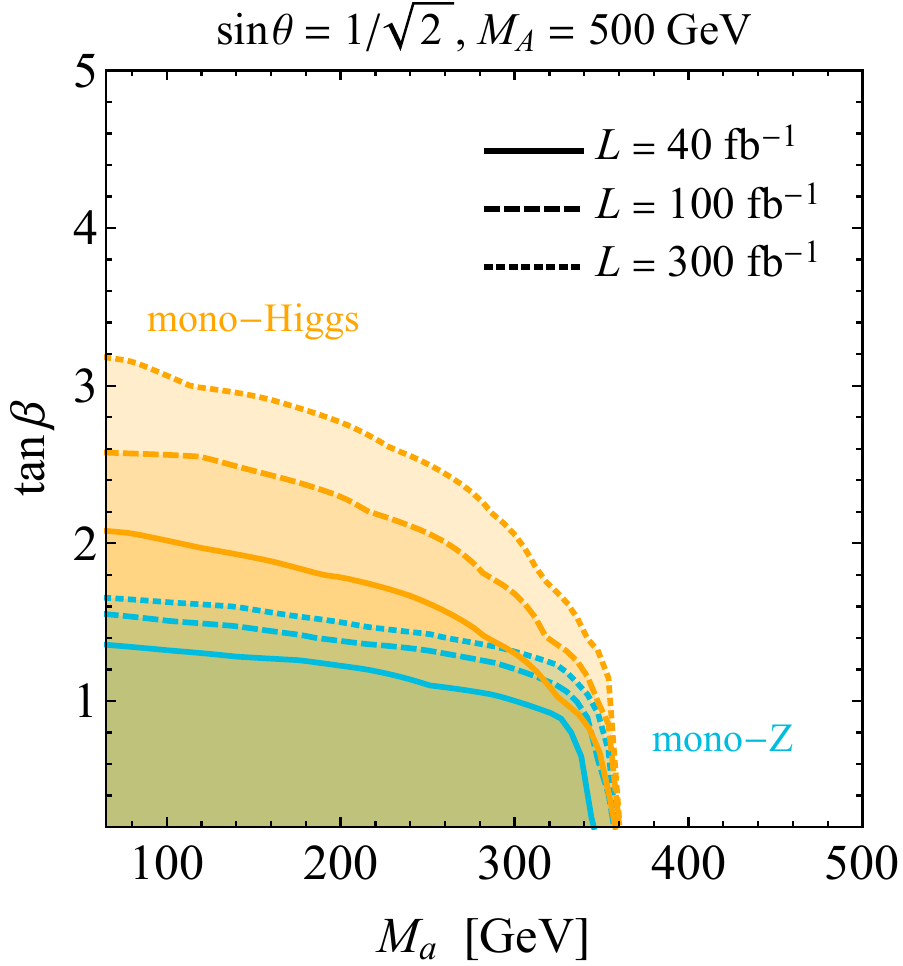}  \qquad 
\includegraphics[width=0.425\textwidth]{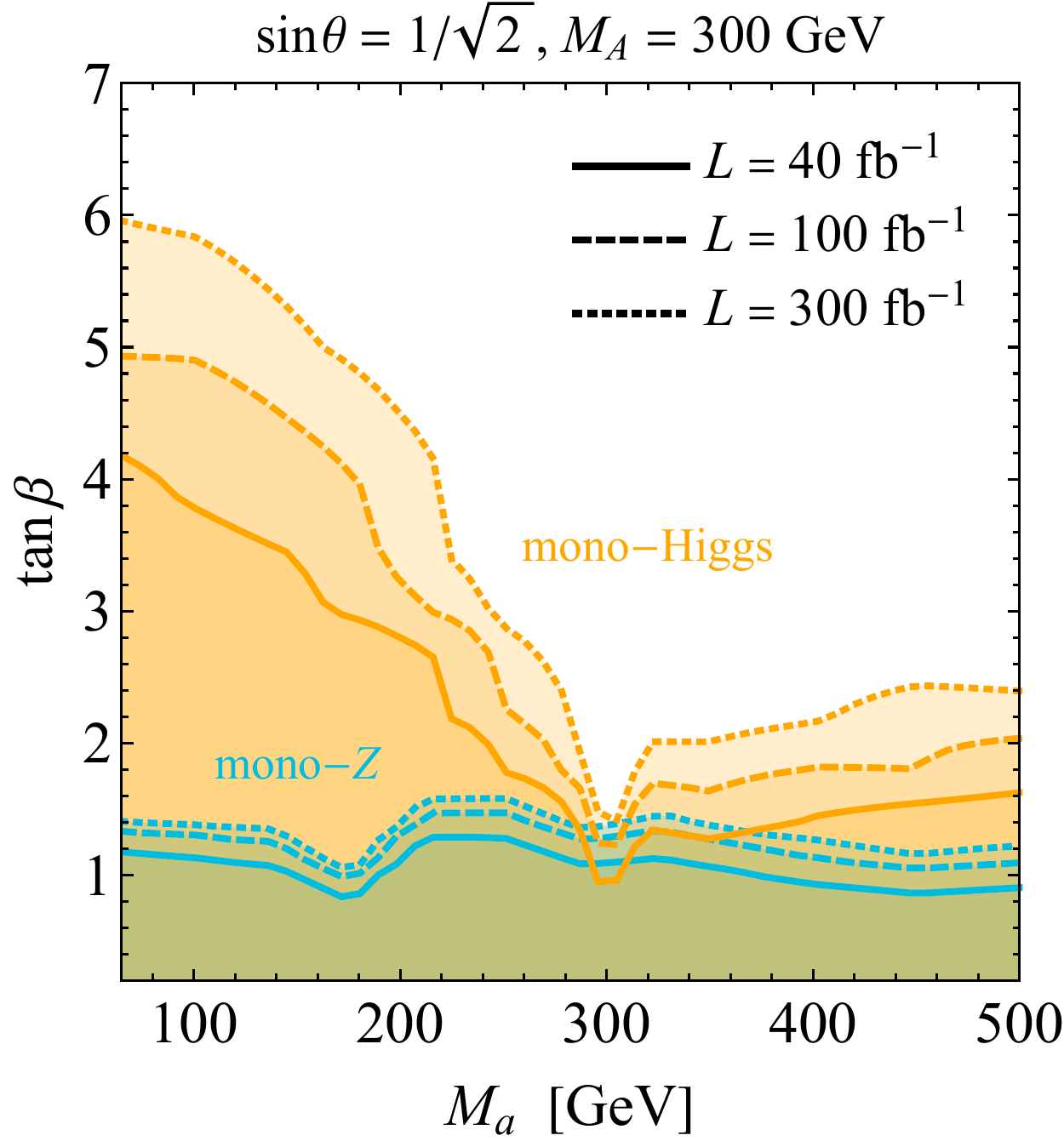} 
\vspace{0mm}
\caption{\label{fig:prospects} 95\%~CL exclusion contours  for our four benchmark scenarios following from hypothetical $Z + E_{T, \rm miss}$~(blue regions) and $h + E_{T, \rm miss}$~(orange regions) searches at 13~TeV LHC energies. The solid, dashed and dotted curves correspond to  integrated luminosities  of $40 \, {\rm fb}^{-1}$, $100 \, {\rm fb}^{-1}$ and $300 \, {\rm fb}^{-1}$, respectively. }
\end{center}
\end{figure}

The future prospects of the mono-$Z$~(blue regions) and mono-Higgs~(orange regions) constraints are illustrated in Figure~\ref{fig:prospects} for  our four benchmark scenarios. We find that by collecting more data the reach of the $Z + E_{T, \rm miss}$ measurements are expected to strengthen, but that the actual improvement depends sensitively on the assumption about the systematic uncertainty on the irreducible SM backgrounds. Assuming a systematic error of 7\%, we observe that the limits on $\tan \beta$ will improve by a mere 10\% when going from $40 \, {\rm fb}^{-1}$ to $300 \, {\rm fb}^{-1}$ of data. In order to further exploit the potential of mono-$Z$ searches, advances in the modelling of $ZZ$ production within the SM would hence be very welcome. 

In contrast to mono-$Z$ searches it turns out that in the case of the $h + E_{T, \rm miss}$ measurements systematic uncertainties will not be a limiting factor even at the end of LHC~Run~II. By increasing the amount of data to $100 \, {\rm fb}^{-1}$ and  $300 \, {\rm fb}^{-1}$, we anticipate that it should be possible to improve the $40 \, {\rm fb}^{-1}$ mono-Higgs limits on $\tan \beta$ by typically 25\% and 50\%, respectively. Notice that larger data sets will be most beneficial in our first and second benchmark scenario in which $\sin \theta$ is small.  In these cases  the resulting $h + E_{T, \rm miss} \, (h \to \gamma \gamma)$  event rates are so low that the sensitivity in the mono-Higgs channel is limited largely by statistics for $40 \, {\rm fb}^{-1}$ of luminosity. 

As explained earlier in Section~\ref{sec:ditau}, we expect that forthcoming searches for  spin-0 resonances in the $\tau^+ \tau^-$ final state should allow to set relevant constraints on $\tan \beta$ in model realisations with a light scalar $H$ of mass $M_H< 2 m_t$. In  the case of our second benchmark scenario this means that it should be possible to test and to exclude the parameter space with $\tan \beta \lesssim {\cal O} (1)$  and $M_a \gtrsim 210 \, {\rm  GeV}$ at LHC Run II. Such an exclusion would indeed be precious,  because as illustrated by the upper right panel of  Figure~\ref{fig:prospects}, this part of the  $M_a$--$\hspace{0.5mm} \tan \beta$ plane is notoriously difficult to  constrain through $E_{T, \rm miss}$ searches. 

\begin{figure}[!t]
\begin{center}
\includegraphics[width=0.425\textwidth]{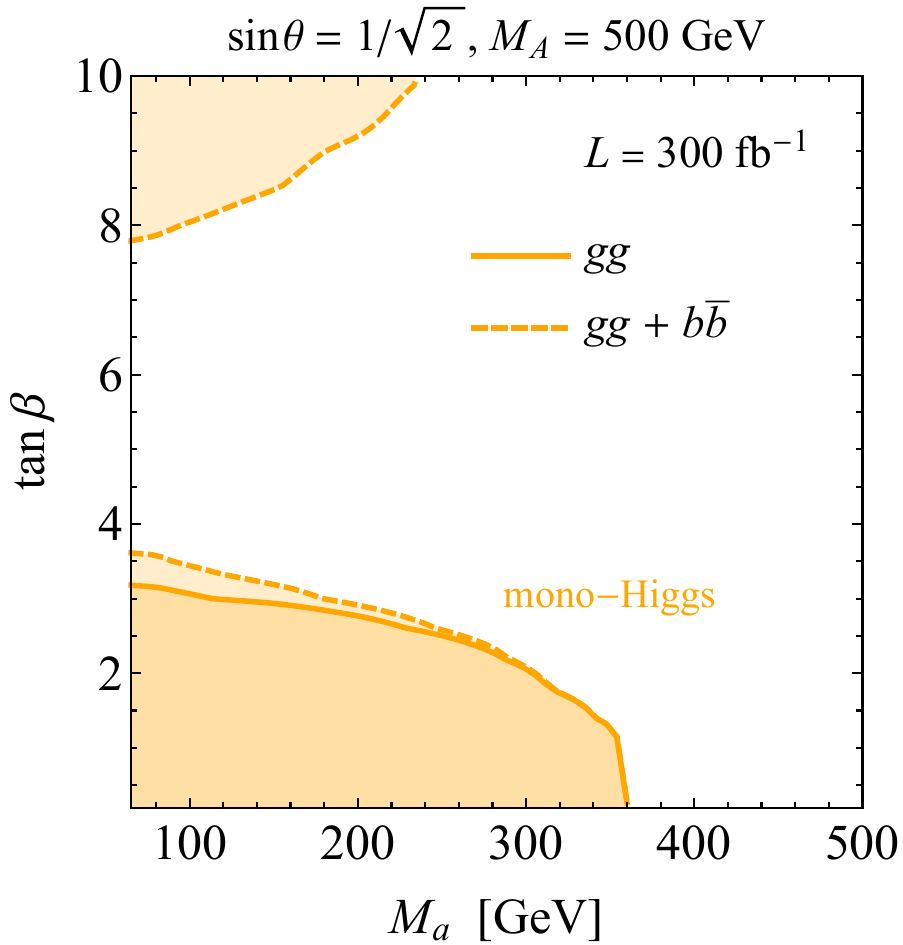}  \qquad 
\includegraphics[width=0.425\textwidth]{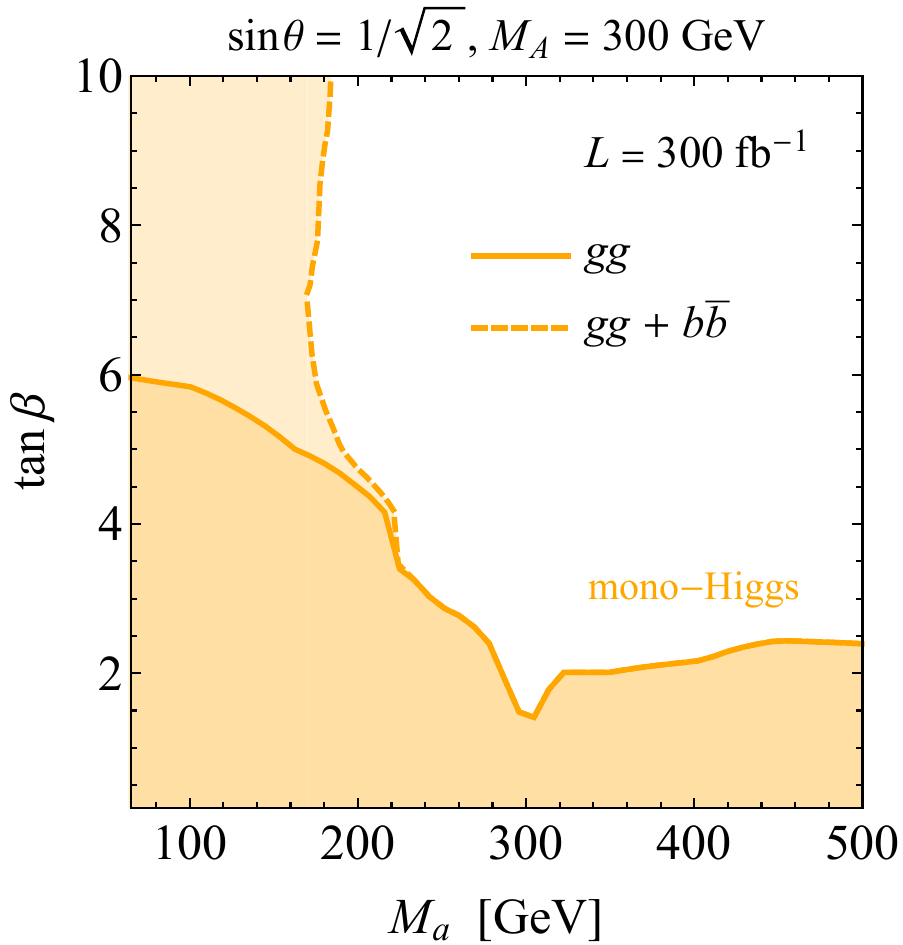} 
\vspace{0mm}
\caption{\label{fig:bbprospects} 95\%~CL exclusion contours  in our third and fourth benchmark scenario that follow from a hypothetical $h + E_{T, \rm miss}$~(orange regions) search with $300 \, {\rm fb}^{-1}$ of 13~TeV data. The solid lines correspond to the limits obtained from $gg$ production alone, while the dashed curves include both the $gg$ and $b \bar b$ initiated channel.}
\end{center}
\end{figure}

Finally, let us comment on an effect already mentioned  briefly in Sections~\ref{sec:monoZ} and~\ref{sec:monohiggs}. In pseudoscalar extensions of the THDM that feature a  $\tan \beta$ enhancement of the bottom-quark Yukawa coupling it is possible in principle to obtain relevant contributions to mono-$X$ signals not only from the $gg \to Z/h + E_{T,\rm miss}$ transitions but also from the $b \bar b \to Z/h + E_{T,\rm miss}$ channels. In Figure~\ref{fig:wetdream} only the model-independent contribution from $gg$ production was taken into account, because the exclusion bounds remain essentially unchanged if also the $b \bar b$-initiated channels are included.

With $300 \, {\rm fb}^{-1}$ of integrated luminosity this situation is however expected to change. Searches for mono-$Z$ signals, for example, should be able to exclude values of $\tan \beta \gtrsim {\cal O} (8)$ for certain ranges of $M_a$  in all four benchmarks. In the third and fourth benchmark scenario there are particularly relevant changes to the projected sensitivity of mono-Higgs searches, as illustrated in~Figure~\ref{fig:bbprospects}. For $M_A = 500 \, {\rm GeV}$ (left panel) we observe that, after including both $gg$ and $b \bar b$ initiated production, model realisations with $\tan \beta \gtrsim 10$ for $M_a \lesssim 220 \, {\rm GeV}$ are excluded. The impact of $b \bar b \to h + E_{T, \rm miss}$ is even more pronounced for a light $A$ with $M_A  = 300 \, {\rm GeV}$~(right panel).  In this case we see that it should be possible to exclude masses $M_a \lesssim 170 \, {\rm GeV}$ for any value of $\tan \beta$. The results displayed in Figure~\ref{fig:bbprospects} have been obtained in the context of a Yukawa sector of type II. Almost identical sensitivities are found in models of type IV, while in pseudoscalar THDM extensions of type I and III bottom-quark initiated contributions are irrelevant, since they are $\tan \beta$ suppressed~$\big($see~(\ref{eq:xicouplings})$\big)$.

\section{Conclusions}
\label{sec:conclusions}

We have proposed a new framework of renormalisable simplified models for dark matter searches at the LHC, namely single-mediator extensions of two Higgs doublet models containing a fermionic dark matter candidate. The mediator can have both scalar or pseudoscalar quantum numbers and all amplitudes are unitary as long as the mediator couplings are perturbative. Constraints from Higgs coupling measurements are averted by mixing the mediator with the heavy scalar or pseudoscalar partners of the Standard Model Higgs. This framework unifies previously established simplified  spin-0 models, while avoiding their shortcomings, and can reproduce several of their features in the appropriate limit. 

In this work we have focused on the case of a pseudoscalar mediator $a$. We have considered the alignment/decoupling limit, in which some of the Higgs partners have masses close to the TeV scale, while either the neutral scalar $H$ or pseudoscalar $A$ is lighter with a mass as low as $300 \, {\rm GeV}$. For the mass of the new pseudoscalar mediator we have considered the range of half the Higgs-boson mass to $500 \, {\rm GeV}$.  These parameter choices are well motivated by Higgs physics, LHC searches for additional spin-0 states, electroweak precision measurements and quark-flavour bounds such as those arising from $B \to X_s \gamma$ and $B$-meson  mixing. Limits on the quartic couplings that arise from perturbativity, unitarity and the requirement that the  total decay widths of $H$ and $A$ are sufficiently small for the  narrow-width approximation to be valid have also been taken into account in our analysis.

By studying the partial decay widths and branching ratios of the spin-0 particles, we have found that the total decay width of the heavier scalar $H$ can be dominated by the $H \to a Z$ channel, while the heavier pseudoscalar $A$ generically decays with large probability through $A \to a h$.  In consequence, the production cross sections for mono-$Z$ and mono-Higgs final states are resonantly enhanced and the obtained limits are competitive with mono-jet searches and even impose the dominant constraints for most of the parameter space at $40 \, {\rm fb}^{-1}$ of 13~TeV LHC data. This surprising result is a consequence of a consistent implementation of the scalar sector and is therefore not predicted by previously considered simplified models (such as the ATLAS/CMS Dark Matter Forum pseudoscalar model). Our findings underline the importance of a complementary approach to searches for dark matter at the LHC and are in qualitative agreement with the conclusions drawn in~\cite{No:2015xqa,Goncalves:2016iyg}. 

We have furthermore  emphasised, that searches for associated production of dark matter with a $t\bar t$ pair will profit from improved statistics unlike the mono-jet search, for which the reach seems systematics limited. We have therefore extrapolated the corresponding constraints to a dataset of $300 \, {\rm  fb}^{-1}$, where  $t \bar t + E_{T, \rm miss}$ searches are  expected  to be more powerful than $j +  E_{T, \rm miss}$ measurements for large parts of the parameter space.

The rich structure of the  two Higgs doublet plus pseudoscalar models has been exemplified by an analysis of four different parameter scenarios. The specific benchmarks have been chosen  to capture different aspects of the mono-$X$ phenomenology that are of interest for future LHC searches.  The results for all scenarios are presented in the form of  $M_a$--$\hspace{0.5mm} \tan \beta$ planes, in which  the  parameter regions that are excluded at 95\% confidence level  by the various $E_{T, \rm miss}$ and non-$E_{T, \rm miss}$ searches have been indicated  (see Figure~\ref{fig:wetdream}). We found that the constraining power of mono-$Z$ and mono-Higgs searches depends sensitively on the mass hierarchies between $M_a$ and $M_A$ or $M_H$, while the sensitivity to the other model parameters such as the amount of mixing in the CP-odd sector is less pronounced. It has also been shown that as a result of the interference of $a$ and $A$ contributions the bounds in the  $M_a$--$\hspace{0.5mm} \tan \beta$ plane that result from  the $j + E_{T, \rm miss}$ and $t \bar t + E_{T, \rm miss}$ channels strengthen above the threshold $M_a \simeq M_A$ in model realisations with a light pseudoscalar~$A$. In addition the reach of the 13 TeV LHC in the mono-$Z$ and mono-Higgs channel has been explored~(see~Figure~\ref{fig:prospects}). While mono-Higgs searches are  found  not to be limited by systematic uncertainties even at the end of LHC~Run~II, in the case of the mono-$Z$ measurements the systematic error can become a limiting factor.  This feature makes the $h + E_{T, \rm miss}$ signal particularly interesting in the context of two Higgs doublet  plus pseudoscalar extensions. 

It has moreover been pointed out that  constraints from di-top resonance searches and flavour observables provide further important handles to test the considered simplified dark matter models. Because the former signature allows to look for neutral spin-0 states with masses above the $t \bar t$ threshold, to which $E_{T, \rm miss}$ searches have only limited access if the dark matter mediators are top-philic, the development of more sophisticated  strategies to search for heavy neutral Higgses  in $t \bar t$ events seems particularly timely. We have also highlighted the possibility to constrain benchmark scenarios featuring  a  light scalar $H$ by forthcoming searches for heavy spin-0 states  in the $\tau^+ \tau^-$ final state, and finally illustrated the impact of bottom-quark initiated production in the case of $h + E_{T, \rm miss}$ (see Figure~\ref{fig:bbprospects}).

To conclude, we stress that meaningful bounds from LHC searches for dark matter can only be extracted if the underlying models are free from theoretical inconsistencies, such as  non-unitary scattering amplitudes or couplings that implicitly violate gauge symmetries. Future ATLAS and CMS analyses of spin-0 mediator scenarios should therefore be based on consistent embeddings of the established ATLAS/CMS Dark Matter Forum  simplified models. For any effort in this direction,  standalone {\tt UFO} implementation of the dark matter models discussed in this article can be obtained from the authors on request. 

\acknowledgments 
We thank all participants of the fourth LHC Dark Matter Working Group public meeting, in particular Nicole Bell, Giorgio Busoni and Jose Miguel No, for interesting discussions. We~are grateful to Stefan Liebler for pointing out the potential relevance of bottom-quark initiated production processes. UH acknowledges partial support by the ERC Consolidator Grant HICCUP (No. 614577) and thanks the CERN Theoretical Physics Department for hospitality.

\end{document}